\documentclass[twocolumn,showpacs,preprintnumbers,amsmath,amssymb,prx,superscriptaddress]{revtex4}
\usepackage{amsfonts}
\usepackage{amsmath}
\usepackage[makeroom]{cancel}
\usepackage[english]{babel}
\usepackage[T1]{fontenc}
\usepackage{times}
\usepackage{mathrsfs}
\usepackage{graphicx}% Include figure files
\usepackage{dcolumn}% Align table columns on decimal point
\usepackage{bm}% bold math
\usepackage{wasysym}
\usepackage[colorlinks,bookmarks=true,citecolor=blue,linkcolor=red,urlcolor=blue]{hyperref}
\usepackage[tight, FIGTOPCAP, hang, raggedright, nooneline]{subfigure}
\usepackage{hyperref}

\begin{document}
\title{Dynamics of Fractionalization in Quantum Spin Liquids}
\author {J.~Knolle*}
\affiliation{T.C.M.~Group, Cavendish Laboratory, J.~J.~Thomson Avenue, Cambridge CB3 0HE, United Kingdom}
\author{D.~L.~Kovrizhin} 
\affiliation{T.C.M.~Group, Cavendish Laboratory, J.~J.~Thomson Avenue, Cambridge CB3 0HE, United Kingdom}
\affiliation{RRC Kurchatov Institute, 1 Kurchatov Square, Moscow 123182, Russia}
\author{J.~T.~Chalker}
\affiliation{Theoretical Physics, Oxford University, 1, Keble Road, Oxford OX1 3NP, United Kingdom}
\author{R.~Moessner}
\affiliation{Max Planck Institute for the Physics of Complex Systems, D-01187 Dresden, Germany}

\begin{abstract}
We present the theory of dynamical spin-response for the Kitaev honeycomb model, obtaining exact results for the structure factor (SF) in gapped and gapless, Abelian and non-Abelian quantum spin-liquid (QSL) phases. We also describe the advances in methodology necessary to compute these results.
The structure factor shows signatures of spin-fractionalization into emergent quasiparticles -- Majorana fermions and fluxes of $Z_2$ gauge field.  In addition to a broad continuum from spin-fractionalization, we find sharp ($\delta$-function) features in the response. These arise in two distinct ways: from excited states containing only (static) fluxes and no (mobile) fermions; and from excited states in which fermions are bound to fluxes. The SF is markedly different in Abelian and non-Abelian QSLs, and bound fermion-flux composites appear only in the non-Abelian phase.
\end{abstract}

\date{\today}

\maketitle

\section{Introduction}
A time-lag of several millennia between the discoveries of ferromagnetic and antiferromagnetic (N\'eel) order, despite their great microscopic similarity, underscores the importance of the availability of experimental probes matching the phenomena in question. The lack of a characteristic macroscopic observable for the N\'eel state has an analogy today in the lack of {\it any} local ground state signatures of topological states of matter, for which the most natural diagnostics -- entanglement entropy or topological degeneracies -- are not readily accessible to present experimental technology. 

The identification of clear-cut experimental signatures is all the more urgent -- after a frustratingly long search following the original proposal of quantum spin-liquid states~\cite{Anderson1973}, there is no longer any shortage of theoretical models exhibiting `topological' quantum spin liquid states~\cite{toric,trirvb,Balents2010}; and in the meantime, several frustrated magnetic materials have been identified as promising candidates to host QSL physics~\cite{Lee2008}.

Perhaps the most natural local diagnostics for spin-liquidity involve the concomitant and characteristic fractionalised excitations above the featureless, long-range entangled, topologically degenerate ground states.  Due to the mismatch between the quantum numbers of such fractionalised excitations on one hand, and the selection rules for standard scattering probes on the other, experiments do not usually couple to a single fractionalised quasiparticle, instead exciting multiple quasiparticles, thereby producing a 
featureless continuum response. 

The dynamical spin response, which can be probed using conventional experimental techniques such as inelastic neutron scattering (INS) and electron spin resonance (ESR), is in principle sensitive to not only the ground state but also to a wide range of excited states, even at zero temperature. As such, it may in particular be sensitive to the presence of fractionalised excitations. Study of the dynamical spin response has proven to be fruitful in applications to one-dimensional systems, where it allowed one to establish a quantitative correspondence between the theoretically predicted correlation functions (obtained exactly using Bethe-ansatz), and the results of inelastic neutron scattering measurements~\cite{Tennant,ruegg}, thereby confirming the presence of $S=1/2$ spinon excitations. 

In the recent Letter~\cite{Knolle2014} we have commenced an analogous programme 
for a two-dimensional quantum spin liquid. The conclusions of Ref.~\cite{Knolle2014,Knolle2014b} were based on an \textit{exact calculation} of the dynamic structure factor for the celebrated 2D Kitaev honeycomb model (KHM); such exact results had thus far mainly been restricted  to one dimension~\cite{Mueller1981,Simons1993,Haldane1993,Pereira2006}.

The Hamiltonian of the Kitaev model is remarkably simple, having only nearest-neighbour exchange. This simplicity has led to a number of theoretical proposals for its realization in condensed matter, and in cold atomic systems~\cite{Jackeli2009,Duan2003}. Materials whose spin and orbital degrees of freedom are strongly entangled in presence of spin-orbit couplings, such as~\{Na,Li\}$_2$IrO$_3$ iridates~\cite{Jackeli2009,cha10,Reuther2011,Coldea2012,Schaffer12,Gegenwart2012}, and more recently~$\alpha$-RuCl$_3$~\cite{Plumb2014,Sears2014,Shankar2014,Sandilands2015,Arnab2015}, are currently the most promising candidates to realise Kitaev physics. Some of these are believed to be in the proximity of a quantum spin liquid state. Remarkably, residual high energy features of these putative QSLs might have already been observed in present systems~\cite{Alpichshev2014,Sandilands2015,Arnab2015}, despite the fact that the latter are known to form a long-range ordered phase.

The KHM represents one of the exceedingly rare instances of a tractable strongly-interacting quantum system in two spatial dimensions~\cite{Kitaev2006}. As a representative of a broader class of QSLs whose emergent degrees of freedom are Majorana fermions and $Z_2$ gauge fluxes, it has become an archetype for a QSL. Despite being formulated a decade ago, it still holds surprises, and is being actively studied, e.g.~in the contexts of the calculations of ground state degeneracy~\cite{Mandal2012}, entanglement entropy~\cite{Yang2007}, transitions between different topological phases~\cite{Feng2007,Xiao-Feng2009,Mukherjee2012}, disorder effects~\cite{Willans2010,Lahtinen2014}, global quench dynamics~\cite{Kells2014,Sengupta2008}, and the effects of doping~\cite{Hyart2012,Vishwanath2012,Trousselet2014,Halasz2014}. There exist a number of integrable generalizations of the model~\cite{Yao2009,Qi2014,Vaezi2014}, as well as its three-dimensional extensions~\cite{Mandal2009,Subhro2014,Kimchi2014,Kimchi2014b,Hermanns2014,Modic2014,Takagi2014,Hermanns2015}.

While the calculation of the time-independent correlators is simple when expressed in appropriate variables~\cite{Baskaran2007}, the calculation of dynamical correlators has turned out to be considerably less straightforward. As noted already in Ref.~\cite{Baskaran2007}, it is possible to map this calculation onto a non-equilibrium problem involving a quantum quench of a local potential, the physics which closely resembles the venerable X-ray edge singularity problem~\cite{Nozieres1969}. 

Here, we aim to provide a \textit{complete theory of the dynamical spin-response in two-dimensional Kitaev QSLs}. We consider various different spin liquids -- both gapped and gapless Abelian as well as gapped non-Abelian. The latter can appear upon breaking time-reversal symmetry, and is of special interest due to proposals of using its non-Abelian excitations for topological quantum computations. For all of these, we provide the numerically exact dynamical structure factor, extending our recent work in Ref.~[\onlinecite{Knolle2014}] 

We find a rich phenomenology, in which each of the considered QSLs appears with distinctive signatures of the emergent fluxes and the Majorana fermions. Some of these properties are rather surprising, such as the appearance of a gap in the response for a gapless QSL, and the existence of a sharp (delta-function) response even for a fractionalised (both gapless and gapped) spin liquid. The explanation of the various features of the response are natural and simple in terms of the fractionalised degrees of freedom, e.g.~involving the gap to a state with a pair of fluxes; or a bound state of the Majorana modes expected for a p-wave superconductor Hamiltonian representing the non-Abelian QSL. 
It would seem hard even to rationalise these phenomena in an alternative language. Therefore, this ensemble of results can be seen as a rather direct validation of the fractionalised picture; given its richness, we do not provide a detailed list of our results in the introduction, and instead devote Section \ref{SecSummary} to a non-technical account of our central results, which has been written with a reader in mind  who is interested in phenomena but not too concerned about technical details. 

Finding the exact solutions presented here has led us to engage in a fair amount of method development. Much of the more technical material included here aims to give a reasonably self-contained account of this. We have in fact developed a number of complementary approaches, both exact (for finite systems based on determinant representation of correlation functions, and in the thermodynamic limit using singular integral equations) and approximate but simple (which we call the adiabatic approximation). It is perhaps worth noting that these should have applicability well beyond the present context. Much of this technical material has been collected into a set of appendices. 

While this paper presents an exact treatment of a particular model QSL, its features  should for the usual reasons be relevant to a much wider range of QSLs, qualitatively and (semi-)quantitatively. Namely, the gap in the response, which originates from the flux gap, the broad continuum due to spin-fractionalization, and the sharp delta-function response due to dynamical rearrangement of Majorana density of states, might hold for QSLs whose low-energy degrees of freedom are heavy fluxes of gauge-field coupled to dispersive fractionalized excitations. These points are discussed as part of our closing outlook section. 
 
The structure of the remainder of this paper is the following. In Section \ref{SecKHM} we introduce 2D honeycomb Kitaev model, and its non-Abelian extension. We summarize our main findings for the dynamical spin-correlators together with the discussion of their qualitative features in Section \ref{SecSummary}. A brief outline of Kitaev's exact solution, is provided for completeness in Section \ref{SecReduction}. In Section \ref{SecSqw} we present details of our calculations of the dynamic structure factor (SF). In Section \ref{SecSqwExact} we outline two complementary exact methods for calculations of the dynamical correlation functions, and in Section \ref{SecSqwApprox} a number of approximate approaches. In Sections \ref{SecResultAbelian} and 
\ref{SecResultnonAbelian} we discuss qualitative features of the structure factor in the whole phase diagram of the extended Kitaev model. The main part of the paper follows with an outlook, Section \ref{SecOutlook}, placing our work in a broader context and outlining directions for further work. 
\begin{figure}[tb]
\begin{centering}
\includegraphics[width=1.0\columnwidth]{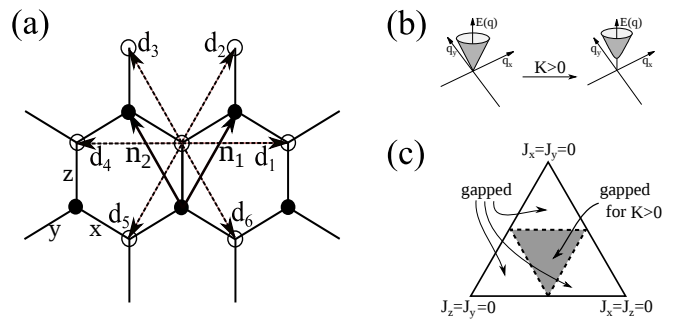}
\par\end{centering}
 \caption{(a) Honeycomb lattice showing bond directions $x,y,z$. Three-spin interactions (with coupling constant K) generate next-nearest-neighbour hopping for Majorana fermions along $\mathbf{d}_i$. (b) Majorana fermion dispersion at the isotropic point  $J_x=J_y=J_z$, which is gapless for $K=0$ (left) and gapped for $K\ne0$ (right). (c) Phase diagram of the extended KHM. For $K\ne0$ the ground state is a gapped non-Abelian QSL in the central triangle (grey), and a gapped Abelian QSL in the outer triangles (white).
\label{lattice}}
\end{figure}

\section{Extended Kitaev model}\label{SecKHM}
The KHM has spin-1/2 degrees of freedom on the sites of a honeycomb lattice. The spins interact via bond-dependent anisotropic Ising exchange $J_a$, where the three directions labeled by $a=x,y,z$ distinguish the three bonds that share a given lattice site, as illustrated in Fig.~\ref{lattice}. In the following we will also discuss an extended KHM, which is obtained from the original model by adding three-spin interactions. The latter is generated by leading order terms in the perturbative expansion in the strength of a small external magnetic field [\onlinecite{Kitaev2006}]. The three-spin interactions break time-reversal symmetry and generate a gap in the spectrum of Majorana fermions, giving rise to non-Abelian excitations~\cite{Lahtinen2011}.

The Hamiltonian of the extended KHM can be written in terms of Pauli matrices $\hat{\sigma}_j^{a}$, and we use the symbol $\langle ij \rangle_a$ to indicate that two nearest neighbour sites (nn) $i,j$ share the same $a$-bond. The extended KHM is obtained by adding next nearest neighbour (nnn) interactions between three spins $\hat\sigma_i^{a} \hat\sigma_j^{c} \hat\sigma_k^{b}$ associated with each pair of bonds $\langle ij \rangle_a$,$\langle jk \rangle_b$ sharing the site $j$, where the direction of the component $c$ is complementary to $a, b$. The Hamiltonian of the extended KHM model is
\begin{equation}
\hat{\tilde{H}}=-\sum_{nn}J_a\hat\sigma_i^a \hat\sigma_j^a 
-K \sum_{nnn} \hat\sigma_i^{a} \hat\sigma_j^{c} \hat\sigma_k^{b}\label{HkitaevThreeSpin}.
\end{equation}
Ground states of the Hamiltonian Eq.~(\ref{HkitaevThreeSpin}) fall into three classes~\cite{Kitaev2006}. For $K=0$ there are two distinct phases, which are gapless and gapped Abelian quantum spin liquids. At non-zero $K$ all of these phases acquire a gap, and the excitations of the formerly gapless state become non-Abelian. In all phases the independent degrees of freedom are static $\mathbb{Z}_2$ gauge fluxes living on the plaquettes of the lattice, and dynamical Majorana fermions defined on the sites. The time-evolution of the Majoranas is generated by the Hamiltonian whose form is fixed by a particular configuration of the $\mathbb{Z}_2$ gauge field.

\begin{figure*}[tb]
\begin{centering}
\includegraphics[width=1.95\columnwidth]{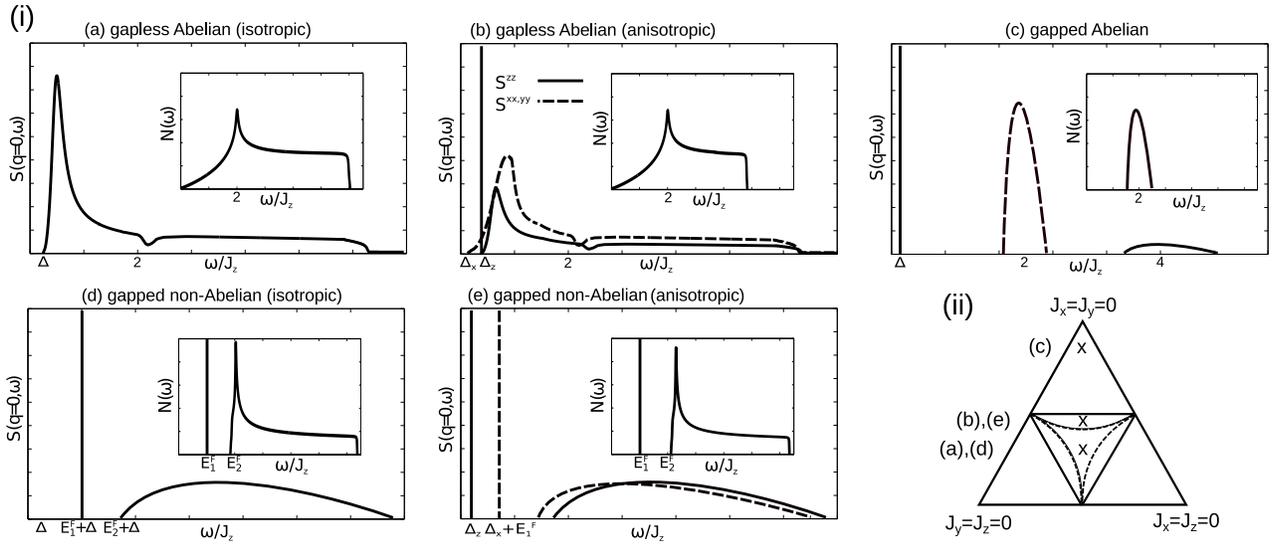}
\par\end{centering}
\caption{(i) Schematic dependence of the structure factor $S^{aa}_{\mathbf{q}=0}(\omega)$ on energy $\omega$ and spin component $a$ for the Abelian and non-Abelian QSL phases of the (extended) KHM. The panels (a) - (e) show behaviour at different representative points in the phase diagram along the line $J_x{=}J_y$, as indicated in (ii), with $K=0$ for (a) - (c) and $K>0$ for (d) and (e). Distinct spin components are denoted by solid ($a=z$) and dashed ($a=x,y$) lines. The insets show the density of states $N(\omega)$ of Majorana fermions and the energy of a Majorana bound state, where this is induced by presence of a flux pair. As discussed in Sec.~\ref{SecSummary}, sharp ($\delta$-function) contributions to $S^{aa}_{\mathbf{q}=0}(\omega)$ appear for the Abelian model in the region of the phase diagram that is unshaded in (ii), but not in the shaded region. An additional sharp component is present in the extended KHM, throughout the non-Abelian phase and in some regions of the Abelian phases. 
\label{Schematic1}}
\end{figure*}

Our central objective is to calculate the dynamical structure factor (SF)
\begin{equation}
\label{DybStrucFact}
S_{{\mathbf q}}^{aa}( \omega)  =   \frac{1}{N}\sum_{ij} 
e^{-i  {\mathbf q} ({\mathbf r}_i-{\mathbf r}_j)} \int_{-\infty}^{\infty} d t e^{i\omega t} S^{aa}_{ij}(t),
\end{equation}
which can be measured in INS and ESR experiments.
It is the Fourier transform in time and space of the dynamical correlation function
\begin{equation}
\label{DynSpinCorrFct}
S^{aa}_{ij} (t)  =  \langle \hat \sigma_i^a(t) \hat \sigma_j^a(0) \rangle,
\end{equation}
where $\hat{\sigma}^{a}_{i}(t)$ is the $a$-th component of spin-operator in Heisenberg representation at time $t$ on site $i$. 

Here we consider zero temperature case, so that the average $\langle ...\rangle$ is taken in the ground state of the Kitaev model. The dynamical structure factor contains extensive information on excitations in the model, as is evident from the Lehmann representation 
\begin{equation}\label{lehmann}
S_{ij}^{aa}(\omega) = \sum_\lambda \langle 0| \hat\sigma_i^a |\lambda \rangle \langle \lambda| \hat\sigma_j^a | 0\rangle \delta(\omega - [E_\lambda - E_0]),
\end{equation}
where $E_\lambda$ is the energy corresponding to an eigenstate $|\lambda\rangle$.

\section{Summary of results}\label{SecSummary}

We find that different representatives of the family of Kitaev Hamiltonians encompass a set of qualitatively different responses, which we depict schematically in Fig.~\ref{Schematic1}. All have in common that the spin correlations are ultra-short ranged, as first noted in Ref.~\cite{Baskaran2007}: the structure factor contains contributions only from on-site and nearest-neighbour correlators, and only those with the same spin-component. This is a consequence of the static nature of the emergent $Z_2$ gauge fluxes. 
As a result, there are no sharp features in reciprocal space -- in itself of course a classic `necessary but not sufficient' diagnostic of spin liquid behaviour -- so that we restrict our plots in Fig.~\ref{Schematic1} to $S_{\mathbf{q}=0}(\omega)$. 

The next and considerably more surprising result is that, in all cases, the dynamical response is gapped, regardless of whether or not the underlying spin liquid phase has an excitation gap. The minimal gap in the response is given by the energy difference between the ground state and the lowest-energy state with a flux pair in adjacent plaquettes. 

One component of the above-gap response is broad in energy and results from Majorana fermion excitations. Its low-energy onset is at the two-flux gap in the phase with gapless Majorana excitations, but is higher in energy in the gapped Majorana phase. As we discuss below, this broad response is due mainly to either single-fermion or two-fermion excitations, depending on spin component and Hamiltonian parameters.

A further striking aspect of the dynamical response is that it includes in some instances sharp ($\delta$-function) components in frequency space, a remarkable feature in view of the fact that all independent quasiparticle excitations of the model are fractionalised and so cannot be created individually by the action of a local operator such as $\sigma_i^a$. In our discussion below, we identify two distinct physical mechanisms by which these sharp contributions arise. One mechanism involves `zero-fermion' transitions, in which only ${\mathbb Z}_2$ fluxes and no Majorana fermions are excited; the other stems from the bound states that are characteristic of vortices in this kind of non-Abelian spin liquid.

While the excitation spectrum of the KHM is independent of the sign of exchange interactions, the ferromagnetic and antiferromagnetic models are clearly distinguished by the $\bf q$-dependence of their response. Viewed in direct space, sign reversal for $J_a$ leaves the on-site correlator $S^{aa}_{ii}$ unchanged but reverses the sign of the nearest neighbour $S^{aa}_{ij}$. In reciprocal space, this sign reversal transfers intensity in a characteristic way between the centre and boundary of the Brillouin zone, as examined in Sec.~\ref{SecResultAbelian}.

Much of the behaviour summarised in Fig.~\ref{Schematic1} can be understood starting from a selection rule for Majorana excitations. We set this out in Sec.~\ref{SubsectionB}, and provide a more detailed discussion of our results for each phase in Sec.~\ref{SubsectionA}. 

\subsection{Selection rules and a dynamical transition}\label{SubsectionB}
The states $|\lambda\rangle$ that contribute to Lehmann expression, Eq.~(\ref{lehmann}), for the dynamical structure factor are ones with non-zero matrix elements $\langle \lambda| \hat\sigma_j^a | 0 \rangle$. Expressed in terms of ${\mathbb Z}_2$ fluxes and Majorana fermions, they obey selection rules which we now discuss.

\begin{figure}[tb]
\begin{centering}
\includegraphics[width=0.5\columnwidth]{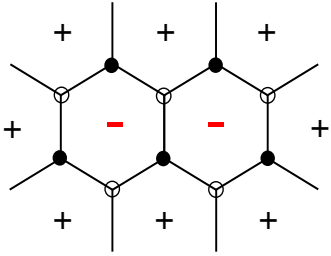}
\par\end{centering}
 \caption{(Colour online). A measurement of a dynamic structure factor leads to a sudden insertion
of a pair of Z 2 gauge-fluxes (shown with minus signs in red).
\label{latticeflux}}
\end{figure}

 The flux selection rule is very simple (see also \cite{Baskaran2007}): the action of $\sigma_j^a $ inserts fluxes through the plaquettes either side of the bond $\langle jk\rangle_a$, as illustrated in Fig.~\ref{latticeflux}. Since the ground state $|0\rangle$ is flux-free and fluxes are static, $|\lambda\rangle$ contains this flux pair and no others.

To introduce the selection rule involving Majorana fermions, consider in the first instance the ferromagnetic Abelian KHM deep in the gapped phase, with $J_z\gg J_x,J_y>0$ and $K=0$. It is then natural to discuss energy eigenstates in the basis of eigenstates of $\hat\sigma_i^z$, and for states in this basis to count the number $N_{\rm z}$ of neighboring pairs of spins  $\langle ij\rangle_z$ 
that are antiparallel. One can define a spin parity operator  $P_{\rm z}\equiv (-1)^{N_{\rm z}}=\prod_j\hat\sigma_j^z$. The Kitaev Hamiltonian commutes with $P_{\rm z}$ and so all energy eigenstates can be chosen to have a definite parity. Moreover, $P_{\rm z}$ is unchanged by the action of a single $\hat\sigma_i^z$ operator, but is reversed by the action of a single $\hat\sigma_i^x$ or  $\hat\sigma_i^y$ operator.
Hence the ground state $|0\rangle$ couples only to states $| \lambda\rangle$ with the {\it opposite} parity for the structure factor components having $a=x,y$, and only to states $|\lambda\rangle$ with {\it same} parity for the component with $a=z$. 

Deep in the gapped phase the lowest energy wave function for Majorana fermions in any flux sector has predominantly $N_{\rm z} =0$, and therefore belongs to the $P_{\rm z}=+1$ sector. The higher energy states with single Majorana fermion excitations consist mostly of one antiparallel spin pair and belong to the $P_{\rm z}=-1$ sector. They form an energy band that is centred on $2J_z$ and has a width set by $J_x$ and $J_y$. 

Components of $S_{ij}^{aa}(\omega)$ with $a=x,y$  therefore arise in this part of the phase diagram only from states $|\lambda\rangle$ that contain odd numbers of fermion excitations. Their main weight is due to single fermion excitations and is concentrated in a band near $\omega=2J_z$. These components also have some weight in higher bands near odd multiples of $2J_z$, but this turns out to be very small.  

Conversely, the component with $a=z$ involves only states with an even number of matter fermion excitations. The lowest in energy of these is simply the unique excited state $|\lambda_0\rangle$ containing no matter fermions and only the added flux pair. Since the matrix element $\langle \lambda_0| \hat\sigma_j^a | 0 \rangle$ involving this state is non-zero, it contributes to $S_{ij}^{aa}(\omega)$ a $\delta$-function with finite weight at frequency $\omega=E_{\lambda_0}-E_0$. Higher energy contributions form bands around even multiples of $2J_z$.

We find a dynamical transition at which this $\delta$-function in the structure factor disappears on moving through the phase diagram. The mechanism is as follows.
Away from the limit $J_z\gg J_x,J_y$, the parities of the ground states in the relevant flux sectors (zero-flux and the three two-flux states that have flux pairs either side of $x$, $y$ or $z$ bonds) are no longer necessarily even. In fact, their relative parities change on a line shown in Fig.~\ref{Schematic1}(ii). All four ground states have the same parity for $J_z\gg J_x,J_y$, but near $J_z=J_x=J_y$ the ground state with a flux pair across a $z$-bond has opposite $P_{\rm z}$ parity to the other three ground states. As a result, there is no $\delta$-function contribution to any component of $S_{ij}^{aa}(\omega)$ in this phase. Note that the boundary on which this dynamical transition occurs is distinct from the previously-known thermodynamic boundary between the gapless and gapped phases, in fact lying within the gapless phase.

Of course, similar arguments can be constructed using parity operators $P_{\rm x}$ and $P_{\rm y}$ based on the other components of spin, leading to the conclusion for the Abelian model that $S_{ij}^{aa}(\omega)$ has a $\delta$-function contribution in a region of the phase diagram where $J_a$ is dominant, but not elsewhere.

There is also a second mechanism that may generate a sharp contribution in the response. It is operative if the spectrum of Majorana fermion excitations that contributes to $\{|\lambda \rangle \}$ includes an isolated level, separate from continuous bands. Let $|\lambda_1\rangle$ be the state containing only a flux pair and this fermion excitation. Provided $\langle \lambda_1| \hat\sigma_j^a | 0 \rangle$ is finite, this state contributes to $S_{ij}^{aa}(\omega)$  a $\delta$-function at frequency $\omega=E_{\lambda_1}-E_0$. This is the case at some values of $J_z \geq J_x = J_y$ in the extended KHM, for $a=x,y$ within both dynamical phases, and for $a=z$ in the dynamical phase that includes the isotropic point. Here the isolated level is a Majorana bound state trapped on the flux pair that is introduced into $| 0 \rangle$ by the action of $\hat\sigma_j^a$. It is known that flux excitations in the non-Abelian phase carry bound states of Majorana fermions, and these spatially localized states appear below the gap of the single-particle Majorana fermion continuum~\cite{Lahtinen2014}.

\subsection{Qualitative features of the response}\label{SubsectionA}

We now discuss more fully the behaviour shown in Fig.~\ref{Schematic1}, where we set $J_x/J_z=J_y/J_z=j$ so that there are only two distinct components of the response: $S^{zz}(\omega)$ and $S^{xx}(\omega)=S^{yy}(\omega)$.

\subsubsection{Abelian QSL}
Schematic results for the structure factor in Kitaev Abelian QSL phases are shown in panels (a) - (c) of Fig.~\ref{Schematic1} (i). 

In the isotropic model (a) $S^{aa}(\omega)$ is non-zero above the energy cost $\Delta$ for introduction of a flux pair. Its dominant weight arises from single Majorana fermion excitations, but a tail continues to higher energy. Although the energy width of the Majorana fermion band determines the extent of the main response, the energy-dependence of the intensity has no simple relation to the magnitude of the Majorana density of states, because the response involves fermion propagation in the presence of a flux pair. It nevertheless reflects features such as the van-Hove singularity.

 In the anisotropic model (b) there are distinct responses $S^{xx}(\omega)$ and $S^{zz}(\omega)$, including different flux gaps, $\Delta_x$ and $\Delta_z$.  Within the gapless phase, both components have non-zero contributions above the respective flux gap. In addition, beyond the dynamical phase boundary $S^{zz}(\omega)$ has a $\delta$-function contribution at $\omega = \Delta_z$. 

In the gapped Abelian phase (c) the $\delta$-function in $S^{zz}(\omega)$ at $\omega = \Delta_z$ persists, but there is an energy gap separating it from the two-fermion continuum around $\omega=4J_z$. By contrast, $S^{xx}(\omega)$ has no $\delta$-function component and is dominated by a single-fermion continuum around $\omega=2J_z$. 

\subsubsection{Non-Abelian QSL}

The response in the non-Abelian phase ($K\ne0$ and $0.5<j<1$) has features that are distinct from the ones which we find for the model with $K=0$. They arise because there are bound states of Majorana fermions associated with flux pairs. 

At the isotropic point (d) this composite flux-fermion bound state manifests itself as a single sharp component in the dynamic structure factor, that would be absent from the corresponding Abelian phase. With anisotropy (e) the energies of sharp contributions to $S^{xx}$ and $S^{zz}$ are the sum of the fermion bound state energy and the two-flux gap, $\Delta_x$ or $\Delta_z$, and therefore unequal. 

\subsection{Broader implications}

The main features of the response described above are robust against, for example, the addition of weak Heisenberg interactions to the Kitaev Hamiltonian, since spin-parity remains a good quantum number. The most important consequence of such additional interactions is that fluxes acquire dynamics. This will broaden the response around $\omega = \Delta$, but we expect that it will remain always gapped, and that distinct contributions to components of $S_{ij}^{aa}$ from states with zero, one and two matter fermion excitations will continue to be identifiable.

\section{Reduction of the spin Hamiltonian to a fermion quadratic form}\label{SecReduction}
The extended KHM model can be solved exactly following the original approach of Kitaev~\cite{Kitaev2006}. We introduce four Majorana fermion species $\hat{c}_{i}$ and $\hat{b}_{i}^{a}$ with $a=x,y,z$ on every lattice site $i$. These fermions obey the anti-commutation relations $\{c_i,c_j\}=2\delta_{ij}$ and $\{\hat{b}_i^{a}, \hat b^{a'}_j\}=2\delta_{ij}\delta_{a a'}$. Spin operators can be represented in terms of $\hat c_{i}$ and $\hat{b}^{a}_i$ as 
\begin{equation}
\hat{\sigma}_i^a=i \hat{c}_i \hat{b}_i^a.
\end{equation} 

Next, we define bond operators $\hat{u}_{ij}=i\hat{b}^{a}_{i}\hat{b}^{a}_j$ with $i,j$ labelling nearest neighbour sites at the ends of bond $a$. In terms of the bond operators $\hat{u}_{ij}$ and the matter fermions $\hat c_{i}$, the Hamiltonian of the extended KHM reads

\begin{align}
\label{majorana}
\hat H   = &  \sum_{\langle ij\rangle_a}  i J_{a}   \hat u _{ij} \hat c_{i} \hat c_{j} +i K \sum_{\langle ij \rangle_{a}, \langle jk \rangle_{b}} \hat u _{ij} \hat u _{jk} \hat c_i \hat c_k.
\end{align}
Bond operators are constants of motion 
with the eigenvalues $u_{ij}= \pm 1$. Thus the Hilbert space in which $\hat H$ acts can be decomposed into `gauge' $|F\rangle$ and `matter' $|M\rangle$ sectors.  Replacing the bond operators by their eigenvalues we arrive at a Hamiltonian which is quadratic in Majorana fermions, and thus can be diagonalised. Note that three-spin interactions give rise to next-nearest-neighbour hopping for the matter fermions.

The Hamiltonian of Eq.~(\ref{majorana}) acts in an enlarged Hilbert space of four Majorana fermions at each site, rather then two-dimensional spin Hilbert space. This redundancy of the Majorana mapping manifests itself in the local $Z_2$ gauge structure, namely the physical properties (including the spectrum) depend on the configurations 
$\{\phi_{\hexagon}\}$ of $Z_2$ fluxes on the plaquettes of the lattice, rather than configuration of bond variables. The flux on each hexagon is given by a product of bond variables $\phi_{\hexagon} = \prod_{\langle ij \rangle \in \hexagon} u_{ij}$. The physical eigenstates $|\Psi_{\rm phys}\rangle=\hat P |\Psi\rangle$ are obtained using a projector to the physical subspace $\hat P=\frac{1}{2}\hat P'\left[1+(-1)^{N_{\chi}}(-1)^{N_{f}}\right]$. Here $\hat P'$ is the sum of all operators which change bond fermion numbers in an inequivalent way~\cite{Yao2009,Loss2011}, and $N_{\chi/f}$ denote bond/matter fermion number operators.

Observables should of course be evaluated using physical eigenstates $|\Psi_{\rm phys}\rangle$, but for the operators which do not change bond fermion number the same result can be obtained by omitting $\hat P$ and employing the unprojected states of the form $|\Psi\rangle=|F\rangle \otimes |M\rangle$ 
(see \cite{Baskaran2007} and Appendix A of Ref.~[\onlinecite{Willans2010}]). In the following we restrict ourselves to observables of this type (note that for large systems complications from finite size effects are negligible~\cite{Zschocke2015}).

For a given configuration of bond variables $\{u_{ij}\}$ the Hamiltonian can conveniently be written in the form
\begin{eqnarray}
\label{HMajorana}
\hat  H   = \frac{i}{2}  
\begin{pmatrix}
\hat c_A &  \hat c_B
\end{pmatrix} 
\begin{pmatrix}
F & M\\
-M^T & -D
\end{pmatrix}
\begin{pmatrix}
\hat c_A\\
\hat c_B
\end{pmatrix}
\end{eqnarray}
with the $N \times N$ matrix $M_{ij}= u_{\langle ij \rangle_a} J_a$ for $N$ unit cells. Here $\hat c_A/\hat c_B$ is shorthand for the $N$-component vectors $\hat c_{A\mathbf{r}}/\hat c_{B\mathbf{r}}$. The next-nearest-neighbour matrices $F_{ij}$ and $D_{ij}$ vanish if $K=0$, but are non-zero at finite $K$, see Eq.~(\ref{majorana}). We note that Eq.~(\ref{HMajorana}) is the most general form of a quadratic Majorana Hamiltonian. 
Instead of dealing with Majorana fermions it is more convenient to work with standard fermions, which can be obtained by combining two Majoranas into a single entity. To this end, we introduce two complex fermion species: bond fermions
\begin{equation}
\hat{\chi}^{\dagger}_{\left\langle i j \right\rangle_a}=\frac{1}{2} (\hat{b}_i^a -i \hat{b}_j^a),
\end{equation}
and matter fermions
\begin{equation}
\hat{f}_{{\mathbf r}}  =  \frac{1}{2} (\hat{c}_{A {\mathbf r}}+i\hat{c}_{B {\mathbf r}}),
\end{equation}
which obey standard anti-commutation relations \cite{Baskaran2007}. Here $A,B$ denote sublattice sites in the unit cell with coordinate~$\mathbf{r}$.
The link variables $\hat{u}_{ij}$ are simply related to the occupation numbers of bond fermions via 
\begin{equation}
\hat u_{ij} =  2 \hat \chi^\dagger_{\langle ij \rangle_a} \hat \chi^{\;}_{\langle ij \rangle_a} - 1.
\end{equation}

Using the shorthand notation $\hat c_A=\hat f^{\dagger}+\hat f$, $\hat c_B=i(\hat f^{\dagger}-\hat f)$ we write the Hamiltonian in terms of complex fermions as 
\begin{eqnarray}
\label{HComplexFermions}
 \hat{H}   =   \frac{1}{2}
\begin{pmatrix}
\hat f^{\dagger} & \hat f
\end{pmatrix} 
\begin{pmatrix}
h & \Delta\\
\Delta^{\dagger} & -h^T
\end{pmatrix}
\begin{pmatrix}
\hat f\\
\hat f^{\dagger}
\end{pmatrix},
\end{eqnarray}
where $h  =   (M+M^T)+i(F-D)$ and $\Delta  =   (M^T-M)+i(F+D)$.
The resulting Hamiltonian has the Bogoliubov-de Gennes form. It is diagonalized using a unitary transformation $T$, see e.g.~Ref.~\cite{ripka}, with $TT^{\dagger}=I$ and
\begin{eqnarray}
\label{UnitaryTrafo}
T
\begin{pmatrix}
h & \Delta\\
\Delta^{\dagger} & -h^T
\end{pmatrix}
T^{\dagger}=
\begin{pmatrix}
E & 0\\
0 & -E
\end{pmatrix}
 \end{eqnarray}
yielding
\begin{eqnarray}
\label{UnitaryTrafo1}
\hat{H}= \sum_{n>0} E_n \hat a_n^{\dagger} \hat a_n  - \frac{1}{2}\sum_{n>0} E_n,
\end{eqnarray}
where $E_n\geq0$ for $n=1\ldots N$ are the eigenvalues, which depend on the flux configuration, $E_n\equiv E_n(\{\phi_{\hexagon}\})$. 
The ground state of the matter fermion Hamiltonian Eq.~(\ref{UnitaryTrafo1}) is defined by $\hat a_i|gs\rangle=0$, with $\hat a_i  = X^*_{ik} \hat f_k + Y^*_{ik} \hat f_k^{\dagger}$. The ground state energy is therefore $E_{gs}=-\frac{1}{2}\sum_{n} E_n$. In order to find the global ground state of the spin Hamiltonian Eq.~(\ref{HkitaevThreeSpin}) one must compare ground state energies $E_{gs} (\{\phi_{\hexagon}\})$ in all flux sectors. Fortunately, due to a theorem by Lieb, we know that the fermionic ground state in a translationally invariant honeycomb lattice is flux-free~\cite{Lieb1994}. 
We denote the ground state of $\hat{H}$ by $|0\rangle = |F_0 \rangle \otimes |M_0 \rangle$, and fix the gauge such that $\hat u_{\langle ij \rangle_a} |F_0\rangle = +1 |F_0\rangle$ for all $\langle ij\rangle_a$.

\subsection{Ground state flux sector}
In the ground state flux sector defined above, the Hamiltonian commutes with translations, and can be block-diagonalized via a Fourier transform $\hat f_{\mathbf{r}}= \frac{1}{\sqrt{N}} \sum_{\mathbf{q}\in \mathrm{BZ}} e^{-i\mathbf{q} \mathbf{r}} \hat f_{\mathbf{q}}$ such that
\begin{eqnarray}
\label{HkitaevThreeSpin2}
 \hat{H}_0 = \sum_{\mathbf{q}\in\mathrm{BZ}} 
\begin{pmatrix}
\hat f_{\mathbf{q}}^{\dagger} &  \hat f_{-\mathbf{q}}
\end{pmatrix} 
\begin{pmatrix}
\xi_{\mathbf{q}} & -\Delta_{\mathbf{q}}\\
 -\Delta_{\mathbf{q}}^{*} & - \xi_{\mathbf{q}}
\end{pmatrix}
\begin{pmatrix}
\hat f_{\mathbf{q}}\\
\hat f_{\mathbf{-q}}^{\dagger}
\end{pmatrix}.
\end{eqnarray}
In this representation $\hat{H}_0$ is equivalent to a BCS Hamiltonian describing a superconductor with a momentum-dependent gap $\Delta_{\mathbf{q}} = -i \text{Im} s_\mathbf{q}-\kappa_\mathbf{q}$ (complex for $K\not=0$), whose quasiparticle dispersion is $\xi_{\mathbf{q}}= \mathrm{Re} s_\mathbf{q}$, where
$s_\mathbf{q}= \sum_{i=0,1,2} J_{\alpha_i} e^{i \mathbf{q} \mathbf{n}_i}$, and $\kappa_{\mathbf{q}}= -4K \sum_{i=1,3,5} \sin{\mathbf{q} \mathbf{d}_i}$.
Here $\alpha_0=z, \alpha_1=x, \alpha_2=y$, the nearest neighbour vectors $\mathbf{n}_0=(0,0)$, $\mathbf{n}_1=(1/2,\sqrt{3}/2)$, $\mathbf{n}_2=(-1/2,\sqrt{3}/2)$, and the six next-nearest neighbour vectors $\mathbf{d}_i$, $i=1\ldots6$ are defined in Fig.~\ref{lattice}~(a).

After writing the expression for the gap in the form $\Delta_{\mathbf{q}}=|\Delta_{\mathbf{q}}|e^{i\phi_{\mathbf{q}}}$, the Hamiltonian $\hat{H}_0$ can be diagonalized by the  Bogoliubov transformation
\begin{eqnarray}
\label{BogTrafo}
 \begin{pmatrix}
\hat f_{\mathbf{q}} \\
\hat f_{-\mathbf{q}}^{\dagger}
\end{pmatrix} =
\begin{pmatrix}
\cos\theta_{\mathbf{q}} & e^{i\phi_{\mathbf{q}}}\sin\theta_{\mathbf{q}}\\
 -e^{-i\phi_{\mathbf{q}}}\sin\theta_{\mathbf{q}} & \cos\theta_{\mathbf{q}}
\end{pmatrix}
\begin{pmatrix}
\hat a_{\mathbf{q}}\\
\hat a_{\mathbf{-q}}^{\dagger}
\end{pmatrix},
\end{eqnarray}
where $\theta_\mathbf{q}$ is fixed by the condition $\tan 2\theta_{\mathbf{q}}=|\Delta_{\mathbf{q}}|/\xi_{\mathbf{q}}$. Defining $E_\mathbf{q}=\xi_\mathbf{q} \cos2\theta_{\mathbf{q}}+|\Delta_{\mathbf{q}}| \sin2\theta_{\mathbf{q}}$ one can write the Hamiltonian in the form
\begin{eqnarray}
\label{Hdiag}
 \hat{H}_0 = \sum_{\mathbf{q}} E_\mathbf{q}(  \hat a_{\mathbf{q}}^{\dagger} \hat a_{\mathbf{q}} -1/2),
\end{eqnarray} 
whose spectrum is given by
\begin{equation}
E_\mathbf{q}=2 \sqrt{\xi_{\mathbf{q}}^2+|\Delta_{\mathbf{q}}|^2}.
\end{equation} 

For $K=0$ the spectrum $E_{\mathbf q}=2|s_{\mathbf q}|$ of fermionic matter excitations $\hat{a}^{\dagger}_{\mathbf{q}} |M_0\rangle$ is gapless if $|J_z|<|J_x|+|J_y|$ (and permutations). At the isotropic point $J_x=J_y=J_z$ there are two Dirac cones positioned at $ \mathbf{Q}=\pm({2 \pi}/{3},-{2 \pi}/{3})$ with a linear energy spectrum $E({\mathbf q})\propto |{\mathbf q}|$ at small energies, see Fig.~\ref{lattice}~(b). In the presence of exchange anisotropy the Dirac cones move in the Brillouin zone, and merge at the transition line (between the gapped and gapless QSLs), so that for $|J_z|>|J_x|+|J_y|$ (and permutations) the spectrum is gapped. The phase diagram of the Kitaev model through the cut in the parameter space defined by $J_x+J_y+J_z=1$ is shown in Fig.~\ref{lattice}~(c).

The Dirac cones of the gapless phase (shown in grey) become gapped for nonzero $K$, see Fig.~\ref{lattice} (b).  The spectrum remains gapless only along the dashed lines in Fig.~\ref{lattice} (c) with quadratic band touching at zero energy. The outer triangles of the phase diagram correspond to gapped Abelian QSLs whose fermionic bands are characterised by a zero Chern number. The particle/hole bands of the formerly gapless phase (central triangle) have Chern numbers $\nu=\pm 1$, and the phase possesses non-Abelian excitations~\cite{Kitaev2006}. 

\section{Calculation of the dynamical structure factor}\label{SecSqw}

In this Section we present several complementary methods which we developed to study the dynamical response in different phases of the (extended)~KHM. Those include two exact methods, as well as a number of approximate approaches. First, we outline the mapping to the quantum quench problem, which is the starting point of our analysis. Second, we present the exact determinant approach, which allows one to study numerically moderately large systems (see details in Appendix A), and may provide a starting point for further analytical investigations. We also discuss another exact approach, based on the solution of an integral equation, which provides results in the thermodynamic limit. We conclude with the discussion of two approximations: the calculation of few-particle contributions to the response, and the adiabatic approach.

\subsection{Quantum quench correspondence}
The calculation of the dynamical spin-response in the (extended) KHM can be mapped onto a local quantum quench problem for Majorana fermions, where a potential is created at time $t=0$, and Majoranas propagate in the presence of this potential at $t>0$, which is similar to a X-ray edge singularity problem. This analogy was noticed by the authors of Ref.~\cite{Baskaran2007}, and the results of our theory have been presented in earlier work~Ref.~\cite{Knolle2014}. Here we outline briefly the main steps of the quantum quench mapping; see Ref.~\cite{Knolle2014} for details.

A general expression for the dynamical structure factor is given by a Fourier transform of a two-time spin-correlation function. The latter can be expressed, following Kitaev, as the matrix element of Majorana fermions and flux operators with respect to the ground state. In this representation the mapping to a quantum quench problem becomes very clear. The spin operator at time $t=0$ in Eq.~(\ref{DybStrucFact}), for example on sublattice `A' given by $\hat \sigma_i^a = i \hat c_i [ \hat \chi_{\langle ij \rangle_a} +\hat{\chi}_{\langle ij \rangle_a}^{\dagger}]$, contains bond fermion operators which exchange their number on the bond $\langle ij \rangle_a$ between zero and one. This corresponds to a change of signs of two fluxes on two adjacent plaquettes sharing the bond, see Fig.~\ref{latticeflux}.  Since the fluxes are static, the spin flip at time $t$ in Eq.~(\ref{DybStrucFact}) must revert these fluxes back to their original state in order to have a non-zero matrix element. This simple selection rule leads to vanishing dynamical spin-correlators beyond nearest-neigbours~\cite{Baskaran2007}, and the ones that remain non-zero are on-site, and nearest-neighbour correlators on the bond $\langle ij \rangle_a$~
\begin{eqnarray}
\label{SpinCorrFct11}
S^{ab}_{ij} (t) & \propto & \delta_{\langle ij \rangle_a}S(t).
\end{eqnarray}

Elimination of the flux degrees of freedom reduces the correlators to an essentially non-interacting form. However, there is a price to pay, as in this representation one is faced with a non-equilibrium problem~\cite{Baskaran2007}, whose physics is closely related to the celebrated X-ray edge singularity (see e.g.~Ref.~\cite{Gogolin2004}). Explicitly, the $z$-components of the correlators, which enter the structure factor read
\begin{align}
\label{SpinCorrFct1}
S^{zz}_{AB} (t) &= - i \langle M_0| e^{i  \hat{H}_0 t} \hat c_{A}  e^{-i \hat H_z t} \hat c_{B}|M_0\rangle,\nonumber  \\
S^{zz}_{AA} (t) &=  \langle M_0| e^{i  \hat{H}_0 t} \hat c_{A}  e^{-i \hat H_z t} \hat c_{A}|M_0\rangle,
\end{align}
and similarly for the $x,y$-components. The Hamiltonian $\hat{H}_z$, which describes the time-evolution of Majorana fermions in the matter sector after the quench
\begin{equation}
\hat{H}_z=\hat{H}_0 + \hat{V},
\end{equation}
differs from $\hat{H}_0$ only in the sign of the nearest- and (in the extended KHM) next-nearest  neighbour Majorana hoppings, as can be seen from the form of a local `quench potential' given by the sum of two contributions $\hat{V}=\hat{V}_z+\hat{V}_K$, where
\begin{align}\label{VflipZ}
\hat{V}_z &=- 2 i J_z \hat c_{A} \hat c_{B},\\
\hat{V}_K &=   2iK [ \hat c_{A}\hat (c_{A \mathbf{d}_5}  -  \hat c_{A \mathbf{d}_6})
+\hat c_{B} (\hat c_{B \mathbf{d}_2} -\hat c_{B \mathbf{d}_3})].
\end{align}
For example the sign of the bond variable $u_{AB}$ in $\hat{H}_z$ is opposite to the one in the ground state. 

This concludes the mapping of the problem of calculating dynamical spin-correlators in Kitaev model to a local potential quantum quench. Despite an obvious similarity of the expressions given above with the ones studied in the X-ray edge problem, we stress that the physics turns out to be quite different, due to the presence of fractionalized quasiparticles.

In the (extended)~KHM the Majorana density of states at small energies either vanishes as zero energy is approached, due to Dirac dispersion, or has a gap, depending on the values of interaction constants $J_a,K$. This low-energy behaviour is in contrast to what appears as an essential ingredient of an X-ray edge singularity problem, namely finite density of states, which lead to power-laws in the response. Related to this is the absence of the $\it{standard}$ Anderson orthogonality catastrophe~\citep{Anderson1967} in our case. For example, deep in the gapless phase, the overlap between two Majorana ground states in different flux sectors (with and without $\hat{V}$) does not vanish in the thermodynamic limit. There is another crucial ingredient in the Kitaev model, that is absent in the X-ray edge singularity problem, and which leads to a new kind of Anderson orthogonality catastrophe. Compared with the standard case, in the Kitaev model only the fermion parity, but not their number, is conserved. We find that this has a dramatic effect on the dynamic correlation functions, and most remarkably, gives rise to a \textit{dynamical phase diagram}: see Fig~\ref{DynamicalPhaseDiagram}.

\subsection{Exact methods}\label{SecSqwExact}
An exact evaluation of the dynamical structure factor starting from the Lehmann representation would amount to a summation of infinite number of multi-particle processes generated by a complete set of states $|\{\lambda_\alpha\}\rangle = \Pi_{\alpha}\hat b_{\lambda_\alpha}^{\dagger}|M_0^F\rangle$ in Eq.~(\ref{Lehmann}). Such a procedure is impractical, and instead we developed two complementary exact approaches of calculating the SF, whose utility varies across the phase diagram.

\subsubsection{Determinant approach for correlation functions}
Rewriting the correlators Eqs.~(\ref{SpinCorrFct1}) in terms of Bogoliubov quasiparticles $\hat a_{q}$ which diagonalize the flux-free Hamiltionian $\hat{H}_0$ [see e.g. Eq.~(\ref{BogNoFlux})] we obtain
\begin{equation}
\label{SpinCorrFctPfaffian}
S^{zz}_{AA/AB}(t) =  e^{i E_0 t}[(X_0^T+ Y_0^T)\hat{M}(X_0^*\pm Y_0^*)]_{00},
\end{equation}
where plus/minus sign corresponds to AA/AB correlators. The main task is the evaluation of the matrix elements of the generic form
\begin{equation}
\label{PfaffianMatrixElement}
M_{ql} (t) = \langle M_0|\hat a_q e^{-i \hat{H}_z t} \hat a_l^{\dagger} |M_0\rangle\,,
\end{equation}
where $H_z$ is a Hermitian operator containing anomalous terms such as $\hat a_q \hat a_k +h.c.$ The latter conserve only the particle number parity, but not their number. 

By representing the Eq.~(\ref{PfaffianMatrixElement}) in terms of a coherent state path integral one can obtain an expression for the matrix elements $M_{ql}$ in terms of Pfaffians 
\begin{equation}
\label{MatrixElementFinal}
 M_{ql} (t) = e^{-i E_{0}^F t }\mathcal{D}_0\{ \mathrm{Pf}[S^{-1}_{\{2N-l,q\}}] - \mathrm{Pf}[S^{-1}]\delta_{ql} \},
\end{equation}
see details and definitions in Appendix~\ref{ExactPfaff}.

This \textit{determinant approach} is exact for finite-size systems in all phases of the (extended) KHM, namely it allows one to obtain the results for time-dependent correlation functions with any desired accuracy at arbitrary times. Similar approaches are used in the studies of quantum quenches in e.g.~quantum Ising model, non-equilibrium Luttinger liquids, as well as in the context of Full Counting Statistics (FCS) \cite{Fagotti2012, Gutman2011, Levitov1993}.

One can obtain a simplified expression for the matrix elements which requires calculation of a single determinant, and an inverse of a $N\times N$ matrix at every time step of the calculations, as shown in  Appendix~\ref{ExactPfaff}. Here we only quote the final result. With the definition of a $N\times N$ matrix
\begin{equation}
\Lambda=\mathcal{Y}^T_Fe^{-i \hat{E}^F t}\mathcal{Y}_F^{*}+\mathcal{X}_F^{\dagger}e^{i \hat{E}^{F} t}\mathcal{X}_F,
\end{equation}
where $\hat{E}^{F}$ is a $N\times N$ diagonal matrix formed from the positive eigenergies $E^{F}_n$ of the Hamiltonian $\hat{H}_z$, and $\mathcal{X},\mathcal{Y}$ are matrices correspond to a product of Bogoliubov transformations, see Appendix~\ref{FluxToFlux}, the matrix elements read
\begin{equation}\label{detexact1}
M_{ql}(t)=\sqrt{\mathrm{Det}[\Lambda(t)]}[\Lambda^{-1}(t)]_{ql}.
\end{equation}
Precise definition of the square root of the determinant can be found at the end of Appendix~\ref{ExactPfaff}.

We note that one can use Eq.~(\ref{detexact1}) in calculations for relatively large systems (we used a laptop to study systems with up to $10^{4}$ spins). In fact, in the gapped (extended) KHM phases it provides essentially numerically exact results for the response because finite size effects are negligibly small even in moderately sized systems. Remarkably, this method also works well near the isotropic point $J_x=J_y=J_z$ because the time-dependent correlation function vanishes quickly with increasing time. This provided us with an independent check of the integral equation approach.

\subsubsection{Integral equation approach}
In a recent Letter we showed that for a KHM it is possible to study the time-dependent correlators exactly in the thermodynamic limit~\cite{Knolle2014}. Here, we outline the main steps of the calculations for completeness.

In the interaction representation, with the time evolution governed by the Hamiltonian $\hat{H}_0$, the local potential $\hat{V}$ plays the role of an interaction. The time evolution of an operator $\hat{A}$ in this representation has the form $\hat A(t)  =  e^{i \hat{H}_0 t}\hat A e^{-i\hat{H}_0 t}$, and the wave-functions evolve under the $\hat{S}$-matrix%
\begin{equation}
\label{Smatrix}
 \hat{\mathcal{S}}(t,0)= e^{i\hat{H}_0t}e^{-i \hat{H}_z t}  =  \mathrm{T}\exp[-i\int_{0}^t d\tau\ \hat{V}(\tau) ],
\end{equation}
where $\mathrm{T}$ is the usual time-ordering, and the nearest-neighbour dynamical correlator defined in Eq.~(\ref{SpinCorrFct1}) assumes the form
\begin{align}
\label{NNCorrFctInteraction}
S_{AB}^{zz}(t) &= -i\langle M_0 |\hat {c}_{A} (t) \mathcal{S}(t,0) \hat {c}_{B}(0) | M_0 \rangle.
\end{align}
The main simplification and the reason why this mapping to an X-ray edge form of the correlator is possible can be traced back to a particularly simple {\it local} form of the impurity potential e.g.~for KHM $\hat{V}_z=-2iJ_z\hat c_{A}\hat c_{B}$, which is clear from the representation in terms of $\hat{f}$-fermions. After introducing the occupation number operator  $\hat{n}_{f}=\hat{f}^{\dagger}\hat{f}$ for the latter, where $\hat{f}^\dagger$ creates a complex matter fermion associated with the bond of the unit cell $\mathbf{r}=0$, the potential can be written as
\begin{equation}
\label{VinBondFermions}
\hat{V}_z(t) = -4 J_z [\hat{n}_f(t)-1/2].
\end{equation}
With these transformations the correlation functions can be reduced to simple expressions
\begin{align}
\label{SpinCorrFctExact4}
S^{zz}_{AB/AA} (t)  &= i[G(t,0)\pm G(0,t)],
\end{align}
where the two Greens functions are given in a standard time-ordered form
\begin{align}
G (t,0)  = -i  \langle  \mathrm{T} [ \hat f(t) \hat f^{\dagger}(0) e^{-i\int_{0}^t d\tau \hat{V}_z(\tau)} ]\rangle,\\
G (0,t)  = -i  \langle  \mathrm{T} [\hat f(0) \hat f^{\dagger}(t) e^{-i\int_{0}^t d\tau \hat{V}_z(\tau)} ]\rangle.
\label{SpinCorrFct1_f}
\end{align}
These expressions for the GFs are similar to the ones which arise in the X-ray edge problem ~\citep{Nozieres1969,Gogolin2004}, and can be evaluated exactly. This was done by us in Ref.~\cite{Knolle2014} using a Dyson equation. The Dyson equation can be solved with the help of methods from the mathematical theory of singular integral equations, see e.g.~a classic book by Muskhelishvili~\cite{mush}, and details of our calculations in the Supplementary Material of Ref.~\cite{Knolle2014}. We checked that our numerical implementation of the \textit{determinant} and \textit{integral equation} approaches produce identical results (see also~\cite{Knolle2014c}).

Note that deep inside the gapless phase, the ground states in the matter sector with and without a flux-pair have a finite overlap in the thermodynamic limit, -- remarkably, there is no Anderson orthogonality catastrophe~\citep{Anderson1967}. This is  due to the fact that the spectrum of matter fermions in the gapless phase is linear (Dirac-like), which leads to a vanishing DOS at small energies, and hence a small number of low-energy excitations which can be generated by an abrupt insertion of the fluxes (whereas in the standard X-ray edge problem the density of states is \textit{finite}). Similarly absent are X-ray edge singularities in the response functions.

\begin{figure}[!t]
\begin{centering}
\includegraphics[width=1.0\columnwidth]{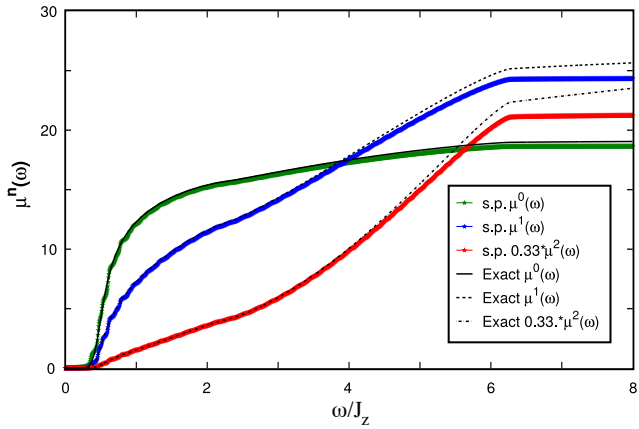}
\par\end{centering}
\caption{Cumulants $\mu^n$ ($n=0,1,2$) of Eq.~(\ref{Cumulant}) are shown for $K=0$ at the isotropic point. Black lines: exact result. Coloured lines: the single particle contribution. Note that, at the frequency of Majorana fermion band edge shifted by the flux gap (i.e.~$\omega=6J_z+\Delta$), the contribution from the single-particle excitations constitute 97.5\% of the response. (Calculation is for a system with 65$\times$65 unit cells. The small oscillations at low energies are due to finite size effects).
\label{SinglePartVersusExactMoments}}
\end{figure}
\begin{figure*}[!t]
\begin{centering}
\includegraphics[width=1.9\columnwidth]{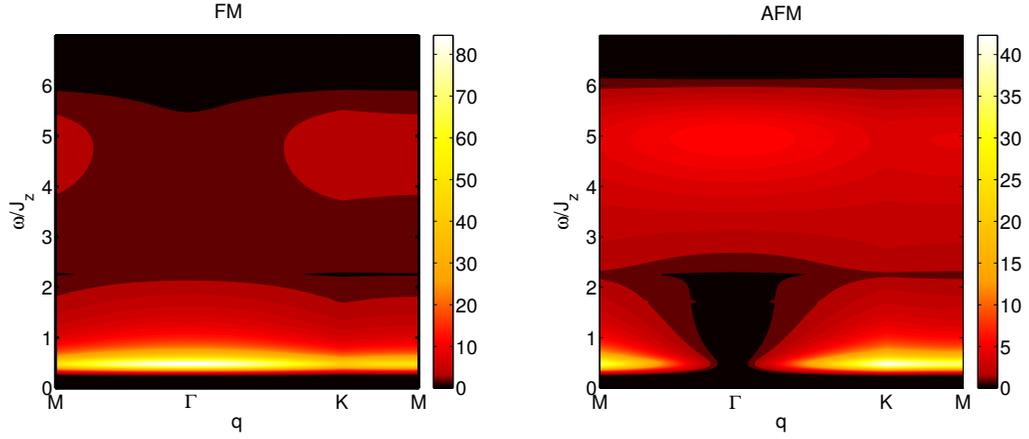}
\par\end{centering}
 \caption{(Color online). 
The dynamic structure factor $S^{aa}(\mathbf{q},\omega)$ for the ferromagnetic (left) and antiferromagnetic (right) KHM at the isotropic point, as  a function of ${\bf q}$ and $\omega$ on the cut $M{-}\Gamma{-}K{-}M$ through the Brillouin zone.
\label{FMvsAFM}}
\end{figure*}
%
%%%%%%%%%%%%%
\subsubsection{Few-particle contributions to the response}\label{SectionFewPart}
Before examining results from the exact solution, it is instructive to look at the Lehmann representation of the dynamical correlation functions in the matter fermion sector. In the remainder we will concentrate on a discussion of the nearest-neighbour correlators $S^{zz}_{AB}(t)$; the on-site correlators $S^{zz}_{AA}(t)$ can be obtained in a similar way. First we define the basis $|\tilde\lambda\rangle$ of many-body eigenstates of the Hamiltonian $\hat{H}_z$ with the corresponding eigenvalues $E^F_{\tilde\lambda}$, where $E_0$ and $E_0^{F}$ are the ground state energies of $\hat{H}_0$ and $\hat{H}_z$ respectively. After insertion of the identity operator $\sum_{\tilde\lambda} |\tilde\lambda\rangle \langle \tilde\lambda|$ into Eq.~(\ref{SpinCorrFct1}) we obtain
\begin{align}
\label{Lehmann}
%\frac
S_{AB}^{zz}(t) = -i\sum_{\tilde\lambda} e^{it(E_0-E^F_{\tilde\lambda})}\langle M_0 |\hat{c}_{A} | \tilde\lambda \rangle \langle \tilde\lambda | \hat{c}_{B} | M_0 \rangle,  
\end{align}
whose Fourier transform gives
\begin{equation}
\label{LehmannFreq}
S_{AB}^{zz}(\omega) = -2 \pi  i \sum_{\tilde\lambda} \langle M_0 |\hat{c}_{A} | \tilde\lambda \rangle \langle \tilde\lambda | \hat{c}_{B} | M_0 \rangle  \delta _{\omega - [E^F_{\tilde\lambda} - E_0]}.
\end{equation}
   
From this representation it is clear that the response vanishes below $\Delta=E_{0}^F-E_0$, which is the energy of the flux gap~\cite{Knolle2014,Tikhonov2010,Tikhonov2011}. 
In a fixed gauge, $\hat{H}_0$ and $\hat{H}_z$ conserve matter fermion parity and the only non-vanishing contributions to Eq.~(\ref{LehmannFreq}) arise from excited states $|\tilde\lambda \rangle$ whose parity is opposite to the ground state $|M_0\rangle$. 
Therefore, the relative matter fermion parity of the ground states with and without fluxes plays a crucial role, and one has two possibilities. In case (I) the ground states of $\hat{H}_0$ and $\hat{H}_z$ have the same parity, in which case the states $|\tilde\lambda \rangle$ must contain an odd number of excitations. In case (II), when the ground states have opposite parity, and $|\tilde\lambda \rangle$ contains an even number of excitations. 
The sector with no excitations makes a special contribution in case (II), because it is just the ground state of the Hamiltonian $\hat{H}_z$. 
Its contribution to $S_{AB}^{zz}(\omega)$ is sharp in frequency, whereas the contributions from the sectors with finite numbers of excitations are broad. As discussed in Sec.~\ref{SecSummary}, this striking distinction between cases I and II gives rise to a \textit{dynamical phase diagram} because the relative parity varies as a function of coupling constants; see also Fig.~\ref{DynamicalPhaseDiagram}.

Using Eq.~(\ref{Lehmann}) one can obtain contributions from different fermion states e.g.~ground state $|M_F\rangle$, single particle states $|\lambda\rangle= \hat b^{\dagger}_{\lambda} |M_F\rangle$, two particle states  $|\lambda\lambda'\rangle= \hat b^{\dagger}_{\lambda} \hat b^{\dagger}_{\lambda'} |M_F\rangle$ etc.,~note the absence of tilde on $\lambda$.

In order to calculate these multiparticle contributions to the response one has to relate fermionic operators in the different flux sectors, as we explain in Appendix \ref{FluxToFlux}. In case (II) (red region in the dynamical phase diagram) the approach must be adapted to the situation in which the relative parity is different; details can be found in Appendix \ref{SectionZeroPart}, where it is shown that a sharp component, which appears exactly at the flux gap, arises from zero particle contribution in the Lehman expansion,   
\begin{equation}
S_{AB}^{zz(0)}(\omega)  \propto  \delta(\omega -\Delta).
\end{equation}

We find (see further discussion at the end of Sec.~\ref{SecSqwApprox}) that the single-particle contribution captures $97.5\%$ of the total weight of the response at the isotropic point of the phase diagram with $K=0$. Moreover, multi-particle contributions become smaller away from this point or at non-zero $K$, and so (depending on relative parities of zero-flux and two-flux ground states) single-fermion or zero- and two-fermion excitations account for nearly all the response.

\subsection{Approximate methods}\label{SecSqwApprox}
\subsubsection{Adiabatic approach}
Due to vanishing density of states in gapless KHM phases a replacement of an abrupt quench of the potential by an adiabatic switching on/off from $-\infty$ to $+\infty$ turns out to be a very good approximation~\cite{Knolle2014} in the green region of the dynamical phase diagram (Fig.~\ref{DynamicalPhaseDiagram}), and this replacement is exact in the low energy limit. In this \textit{adiabatic approach} the potential generated by a flip of a flux-pair is switched slowly in time, thus the $S$-matrix $\mathcal{S}(t,0)$ can be replaced by $\mathcal{S}(\infty,-\infty)$, see e.g.~\citep{AGD}, and the correlator assumes the form
\begin{multline}
\label{NNCorrFctAdiabatic}
S_{AB}^{zz,\text{ad}}(t) = -i\langle M_0 |T[\hat{c}_{A} (t)\hat{c}_{B}(0)\mathcal{S}(\infty,-\infty)]| M_0 \rangle \\ 
  = -ie^{it E_0}\langle M_F |\hat{c}_{A} e^{-i\hat{H}_z t } \hat{c}_{B}| M_F \rangle
\end{multline}
with the only difference between Eqs.~(\ref{NNCorrFctAdiabatic}) and (\ref{SpinCorrFct1}) being that the ground state $|M_0\rangle$ has been replaced by $|M_F\rangle$. 
This dramatically simplifies the calculations (because the integral equation can now be solved by a Fourier transform), and one arrives at an expression for the structure factor in this approximation
\begin{equation}
\label{StructureFactor1partAdiabatic}
S_{\mathbf{q}=0}^{zz,\text{ad}}(\omega)  =  8 \pi \sum_{\lambda}  |X_{\lambda 0}|^2\delta_{\omega -\left[E_{\lambda}^F+\Delta\right]},
\end{equation}
where $X$ is a Bogoliubov matrix, see Appendix \ref{FluxToFlux}.

\subsubsection{Lehmann representation and single-particle contributions}
Another quantitatively good approximation can be derived from the Lehmann representation Eq.~(\ref{LehmannFreq}).
It holds in case (I) (green region in Fig.~\ref{DynamicalPhaseDiagram} (a)) where both ground states (with and without fluxes) have the same parity, so that only the states with odd numbers of fermion excitations parities contribute. Taking the $|\lambda \rangle = \hat{b}_{\lambda}^{\dagger} |M_0^F\rangle$ we find the single-particle contribution to the structure factor
\begin{multline}
\label{StructureFactor1part}
S_{\mathbf{q}=0}^{zz(1)}(\omega)  =  8 \pi  |\langle M_F|M_0 \rangle|^2  \\ \times\sum_{\lambda}|[\mathcal{X}^{-1\dagger}X]_{\lambda0}|^2\delta_{\omega -[ E_{\lambda}^F + \Delta]},
\end{multline}
see Appendix \ref{SectionSinglePartC} for details and definitions. Similarly, one can obtain two-particle contributions, see e.g.~Fig.~[\ref{StructureFacThreeSpinK01Jxy015n45}].

Comparison between cumulants 
\begin{align}
\mu^{[n]}(\omega)=\int_0^{\omega} d\Omega\ \Omega^n S_{\mathbf{q}=0}(\Omega) 
\label{Cumulant}
\end{align}
of the exact solution with those of the single-particle response is presented in Fig.~\ref{SinglePartVersusExactMoments}. The agreement is remarkable, which is due to the fact that only a small number of fermionic excitations is generated by the quench. The absence of orthogonality catastrophe makes the single particle contribution, Eq.~(\ref{StructureFactor1part}), a quantitatively good approximation accounting for 97.5\% of the total intensity.  Note that, above the single-particle Majorana band edge adjusted by the flux gap $\omega=6J_z+\Delta$, the single-particle cumulants $\mu^{[n]}$ do not depend on $\omega$. In contrast, the cumulants of the exact solution are frequency dependent due to many-particle processes (whose contribution to the response is very small).

\section{Results for the structure factor}

In this Section we present a selection of the results for the structure factor in support of the schematic pictures shown in Fig.~\ref{Schematic1}. The results here supplement those in Fig.~2 of our earlier paper \cite{Knolle2014}.

\subsection{Structure factor of the gapless Abelian QSL}\label{SecResultAbelian} 

An overview of the wave-vector and frequency dependence of the structure factor, and a comparison between the ferromagnetic and antiferromagnetic models, is shown in Fig.~\ref{FMvsAFM} for the isotropic point of the KHM. The main features of the $\omega$ dependence are relatively insensitive to $\bf q$ and to the sign of the exchange interactions. However, there is a striking shift of intensity from the center of the Brillouin zone in the ferromagnetic case to the edge of the Brillouin zone in the anti-ferromagnetic case, because of the change in sign of the nearest-neigbour correlator $S_{ij}^{aa}(\omega)$, as discussed in Sec.~\ref{SecSummary}.

\begin{figure}[tb]
\begin{centering}
\includegraphics[width=1\columnwidth]{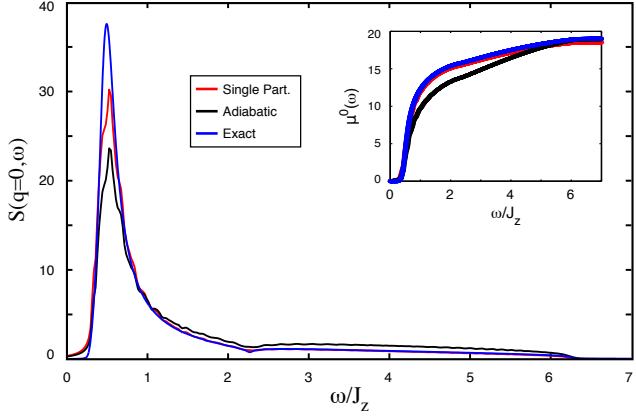}
\par\end{centering}
\caption{Dependence of $S^{zz}_{\mathbf{q}=0}(\omega)$ on $\omega$ at the isotropic point of KHM. Blue line (highest peak) is a numerically exact result.
Red line (middle peak), single particle contribution. Black line (lowest peak), the result of adiabatic  approximation. The blue line is obtained in the thermodynamic limit via the integral equation approach; the red and the black lines are obtained for a system with 65$\times$65 unit cells, with energy averaging over a window of width $0.025J_z$ to remove finite size effects.
Comparison between cumulants calculated within the same approaches is shown in the inset.
\label{SinglPartVsAdiabatic}}
\end{figure}
\begin{figure}[tb]
\begin{centering}
\includegraphics[width=1.0\columnwidth]{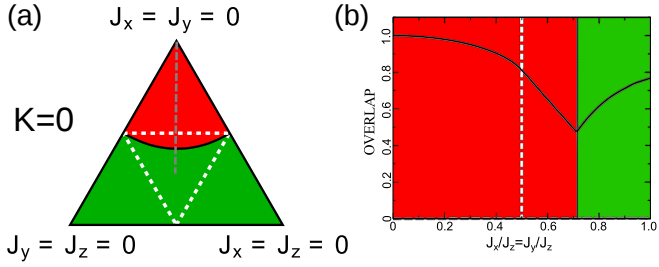}
\par\end{centering}
 \caption{(a) Dynamical phase diagram of the KHM model. The structure factor $S^{zz}(\omega)$ contains a $\delta$-function contribution in the red region of panel (a), but not in the green region. Thermodynamic phase boundaries are indicated by white dashed lines. (b) Overlaps between Majorana fermion ground states from different flux sectors on the line $J_x=J_y$ (see Appendix.~\ref{SectionZeroPart} for notation). The weight of the $\delta$-function, which is proportional to the overlap $|\langle M_F^{xy}|M_0 \rangle|^2$, is shown in the red region. At the dynamical phase transition, $J_x/J_z=J_y/J_z\approx 0.71$, the overlap drops to zero. The alternative overlap $|\langle M_F|M_0 \rangle|^2$, shown in the green region, is finite where $S^{zz}(\omega)$ has no $\delta$-function contribution.
\label{DynamicalPhaseDiagram}}
\end{figure}

\begin{figure}[!t]
\begin{centering}
\includegraphics[width=1.0\columnwidth]{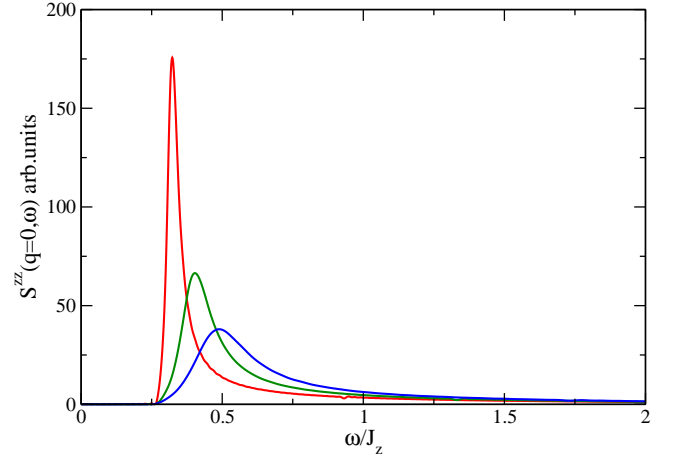}
\par\end{centering}
\caption{$S^{zz}_{\mathbf{q}=0}(\omega)$ component of the dynamical structure factor in the gapless phase (green in Fig.~\ref{DynamicalPhaseDiagram}) for $J_z=1$. Red, green, and blues curves: f $J_x=J_y=0.8,0.9,1.0$. Note that the response diverges at the threshold on approach to the dynamical phase boundary. A similar divergence appears in the calculation of the the adiabatic approximation, 
Eq.~(\ref{StructureFactor1partAdiabatic}). 
\label{divergence}}
\end{figure}

\begin{figure}[!t]
\begin{centering}
\includegraphics[width=1.0\columnwidth]{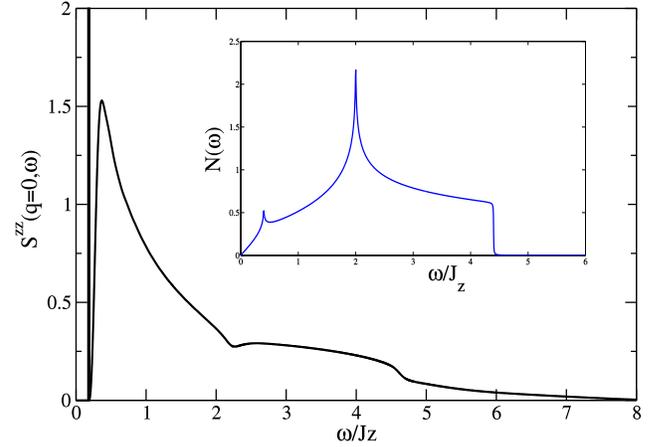}
\par\end{centering}
\caption{$S^{zz}_{\mathbf{q}=0}(\omega)$ component of the dynamical structure factor in the gapless intermediate phase (red in Fig.~\ref{DynamicalPhaseDiagram}), for $J_x=J_y=0.6$, $J_z=1$. Note a $\delta$-function contribution to the response at the energy of the flux-gap. The broad component of the structure factor shows significant multi-particle weight compared to other phases.
\label{ResponseDynamicalPhase}}
\end{figure}

The frequency dependence of $S^{aa}(\mathbf{q},\omega)$ with ${\bf q}=0$ is shown in Fig.~\ref{SinglPartVsAdiabatic} for the ferromagnetic model at the same point in the phase diagram. The main features set out in Sec.~\ref{SecSummary} are apparent: response is non-zero only above the two-flux energy gap, and the dominant contribution extends over the energy width of the Majorana fermion band. Fig.~\ref{SinglPartVsAdiabatic} also demonstrates that single Majorana fermion excitations account for the majority of the response, and that our adiabatic approximation [scaled using the sum rule Eq.~(\ref{SumRules})] captures the behaviour quite accurately.

Evidence for the dynamical phase transition, which is discussed in detail in Sec.~\ref{SecSummary}, is presented in Fig.~\ref{DynamicalPhaseDiagram}. The structure factor $S^{zz}(\omega)$ includes a $\delta$-function contribution in the indicated region of the phase diagram, with a weight that drops discontinuously to zero at the transition. On approaching the transition from the opposite side, a broad peak above the flux gap (see Fig.~\ref{SinglPartVsAdiabatic}) sharpens and shifts to lower energies, see Fig.~\ref{divergence}. The continuum response has a divergence at the flux gap energy precisely at the transition. After crossing the transition, this sharp peak splits into a $\delta$-function, and a finite continuum response as shown in Fig.~\ref{ResponseDynamicalPhase}.

\subsection{Structure factor of the extended KHM ($K\not=0$)}\label{SecResultnonAbelian}
\begin{figure}[t!]
\begin{centering}
\includegraphics[width=0.95\columnwidth]{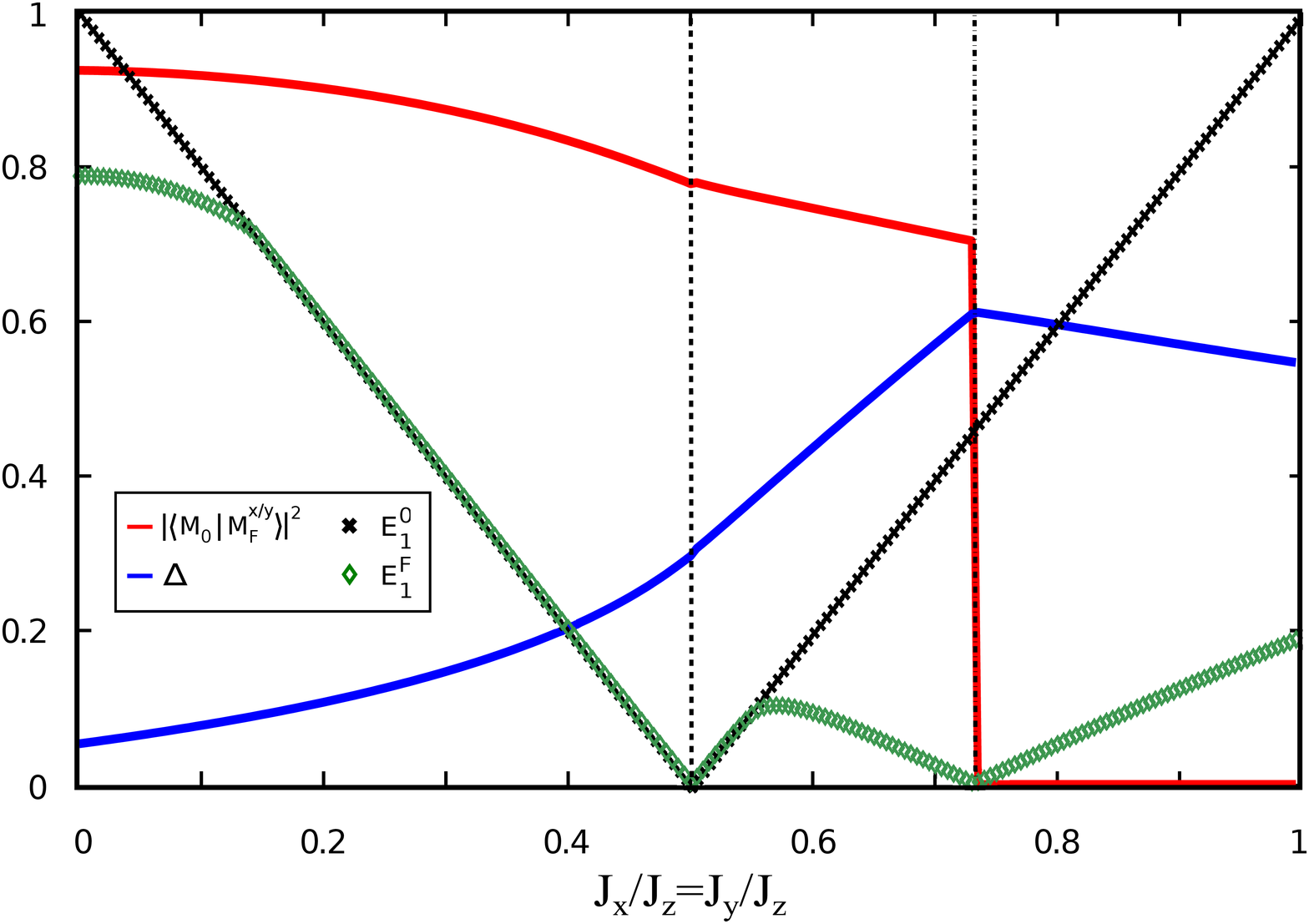}
\par\end{centering}
 \caption{Spectral features on the line $j{=}J_x/J_z{=}J_y/J_z$ for $K=0.1J_z$, shown in order to highlight the signatures of the thermodynamic phase transition between Abelian and non-Abelian phases at $j=0.5$ and the dynamical phase transition at $j\simeq 0.73$. Red line: overlap $|\langle M_0|M_F^{xy}\rangle|$; blue line: the flux gap $\Delta$. Both the overlap and the flux gap are continuous at $j=0.5$, but their derivatives are discontinuous, signalling the thermodynamic transition~\cite{Xiao-Feng2009}, while at $j\simeq 0.73$ the overlap drops abruptly to zero, indicating the dynamical phase transition. Black crosses: dependence of the energy of the lowest excited state (the band edge) for a translationally invariant system (black crosses); green diamonds: energy of the lowest excited state for a system with two fluxes (green diamonds). For the values of $j$ where these energies are different, the fluxes support a static (as opposed to a dynamically generated) fermion bound state. 
\label{OverlapFluxGapK01n52}}
\end{figure}

At a finite value of the three-spin interaction $K$, the next-nearest neighbour hopping gaps out the Dirac cones (see Fig.~\ref{lattice}), and the resulting QSL state hosts non-Abelian excitations. This can be seen in the representation of the Hamiltonain in terms of spinless fermions, Eq.~(\ref{HkitaevThreeSpin2}). These interactions break time-reversal symmetry, and the ``superconducting gap'' in Eq.~(\ref{HkitaevThreeSpin2}) acquires a phase. The Hamiltonian in the translationally invariant system (in the absence of fluxes) is identical to a Hamiltonian describing $p_x+i p_y$ superconductor, which can support bound states in vortex cores. More specifically, an isolated half-vortex, equivalent to a ${\mathbb Z}_2$ flux in the KHM, carries a zero energy state~\cite{Ivanov2001,Kitaev2006,Kells2010}. 

When evaluating the dynamical structure factor we are concerned with fermion states in the presence of a {\em pair} of ${\mathbb Z}_2$ fluxes, induced by the action of $\sigma^a_j$ on the flux-free found state. Since the two fluxes are in adjacent plaquettes, the zero energy modes that would be associated with each one if they were isolated are hybridised, forming a more conventional bound state at finite energy, or merging with the continuum for some values of $J_a$. The behaviour of this bound state and some other features of the Majorana fermion spectrum in the extended KHM are illustrated in Fig.~\ref{OverlapFluxGapK01n52}.

\subsection{Fermionic bound state signatures in the structure factor}
\begin{figure}[tb]
\begin{centering}
\includegraphics[width=1\columnwidth]{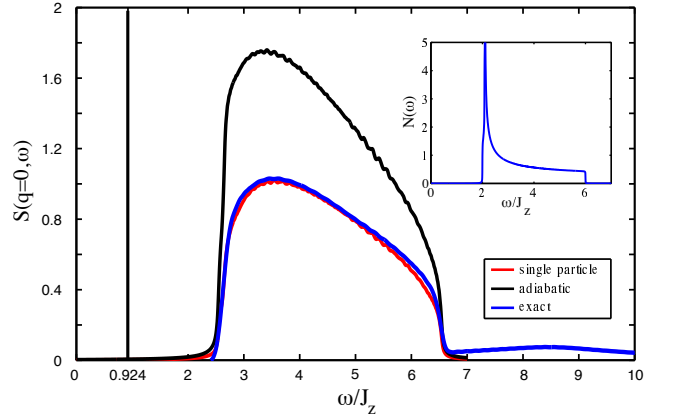}
\par\end{centering}
\caption{Dynamic structure factor $S^{aa}_{\mathbf q}(\omega)$ at $J_x{=}J_y{=}J_z$ and $K=0.1J_z$ as calculated from the exact {\it determinant approach} (blue). In addition we show the single-particle contribution (red) and the adiabatic approximation (black, with total intensity fixed by the sum rule). Inset: Majorana fermion density of states $N(\omega)$ of the flux-free sector. Note a $\delta$-function contribution to $S^{aa}_{\mathbf q}(\omega)$ from the localized Majorana state bound to a flux pair. Its energy, $\Delta+E_1^F=0.924 J_z$, is below the single particle gap. The calculations are done for a system of $56 \times 56$ unit cells, using an energy broadening of $0.025 J_z$ to reduce finite-size effects. 
 \label{ThreeSpinStructurFJz1Jx1Jy1}}
\end{figure}
In Fig.~\ref{ThreeSpinStructurFJz1Jx1Jy1} we present results for the dynamic structure factor at the isotropic point in the non-Abelian phase. The main features of the response are a $\delta$-function component due to a Majorana fermion level bound to a flux pair, and a broad component from excitations to the fermion continuum. The sharp feature is at an energy $\omega=\Delta + E_1^F = 0.924J_z$ which is the sum of the two-flux gap $\Delta=E_{0}^F-E_0 \simeq 0.545J_z$ and the fermion bound state energy $E_1^F$, while the onset of the broad feature is at $\Delta+E^F_2$, where $E^F_2=2.002J_z$ is the energy of the fermion band edge.

\begin{figure}[tb]
\begin{centering}
\includegraphics[width=1.0\columnwidth]{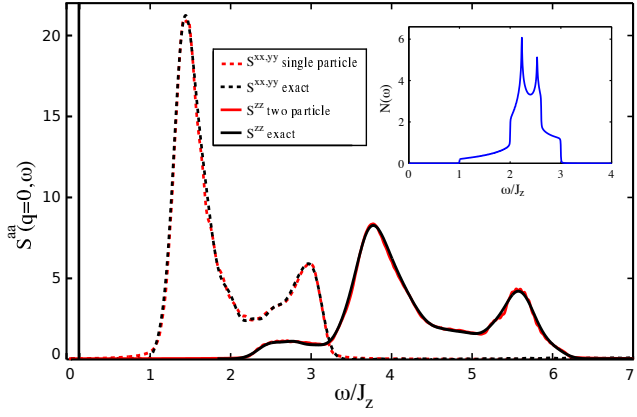}
\par\end{centering}
 \caption{Dynamic structure factor $S^{aa}(\omega)$ for inequivalent components $a{=}z,x$ in the Abelian anisotropic QSL at $J_{x}{=}J_{y}{=}0.25 J_{z}$ and $K=0.1J_z$.
Inset: Majorana fermion density of states $N(\omega)$ in flux-free sector. 
The $zz$-correlator has a $\delta$-function contribution at the two-flux energy, arising from a `zero-fermion' excitation, and  also a weak two-fermion contribution.
The main contribution to the $xx$-correlator is from single fermion excitations.
Calculations are done for a system of size $56 \times 56$ unit cells using a broadening $0.025 J_z$ to reduce finite size effects. 
\label{StructureFacThreeSpinK01Jxy015n45}}
\end{figure}

In Fig.~\ref{StructureFacThreeSpinK01Jxy015n45} we show the dynamic  structure factor $S^{aa}(\omega)$ at a point in the Abelian phase of the extended KHM. It displays some of the main features introduced in Sec.~\ref{SecSummary}: a distinctive difference between $S^{xx}(\omega)$, dominated by single-fermion excitations, and $S^{zz}(\omega)$, which has a sharp zero-fermion contribution at the energy $\Delta=0.0905 J_z$ of the flux-gap, and a small two-fermion band, with an onset at $\Delta+2 E_1^F$, where $E_1^F$ is the lowest fermion excitation energy, lying at the band edge since there is no fermion bound state for this choice of interaction strengths.

\begin{figure}[t!]
\begin{centering}
\includegraphics[width=1.0\columnwidth]{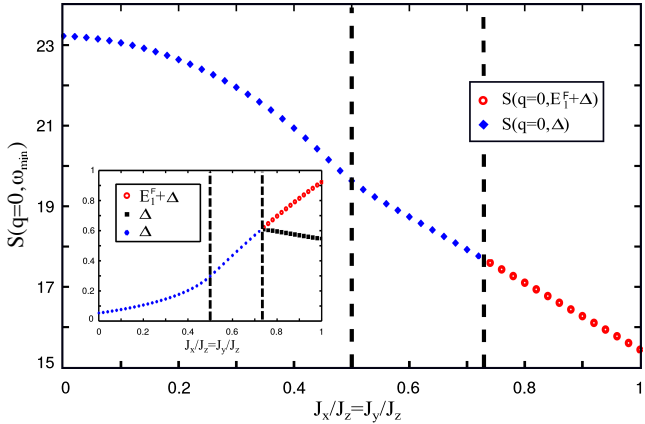}
\par\end{centering}
 \caption{Dependence of the weight of $\delta$-function contribution to $S^{zz}(\omega)$ on $j=J_x/J_z=J_y/J_z$ for $K=0.1J_z$. Inset: dependence of the energy $\omega_{\mathrm{min}}$ of the $\delta$-function contribution and the two-flux gap on $j$. The thermodynamic and dynamical transitions occur at $j=0.5$ and $j=0.73$ respectively. Weights and energies vary continuously across the transitions, although the origin of the  $\delta$-function is different in the two dynamical phases.
\label{DeltWeighJ}}
\end{figure}
We examine in Fig.~\ref{DeltWeighJ} the evolution across the phase diagram of the energy and weight of the $\delta$-function contribution to $S^{aa}(\omega)$ in the extended KHM. This arises through different mechanisms in the two dynamical phases. For $j<0.73$ it is due to a transition to a state with a flux pair but no fermion excitations, and is visible only in the component $a=z$. For $0.73<j\leq1$ it arises from a transition to a state with a flux pair and a fermion excitation in a bound state, and appears in all components $a$. It is notable that evolution is continuous across the dynamical transition and also across the thermodynamic transition at $j=0.5$.

\section{Outlook}\label{SecOutlook}
We have presented a complete and exact calculation of the dynamical structure factor of the various quantum spin liquid phases of the Kitaev honeycomb model, with results summarised in Section~\ref{SecSummary}. These contain both general features indicating the presence of fractionalised excitations, as well as a sufficient amount of detail reflecting the particular QSL phase to be useful as identifying diagnostics.

A question which follows almost reflexively concerns the applicability of such a set of results to models away from special points of exactitude, or indeed actual materials. A case in point is the kind of Heisenberg-Kitaev $J_1-J_2-J_3-\Gamma$ model which has been adduced to account for properties of the Ir-Kitaev materials in 2D. 

Here, we expect important gross features to be robust, as the mechanisms underpinning these phenomena are not predicated on integrability. For instance, the emergent fluxes will continue to be natural variables to describe the spin liquid, even if they are no longer immobile or entirely absent from the ground state. A suppression of low-energy scattering due to the selection rule on the fluxes should therefore persist. Similarly, for the sharp delta-function features, both the ingredients underpinning their appearance survive departure from integrability. Given these two -- parity selection rules, and the presence of Majorana bound states in the case of the non-Abelian QSL -- can further be distinguished by considering different components of the structure factor in the anisotropic spin response, this may be a particularly promising way for detecting a phase with non-Abelian anyons. Regarding the latter, however, it is worth bearing in mind that there is  the possibility of the bound state energy being increased to the extent that it will merge with the single-particle continuum.

Our results indicate that the dynamical structure factor allows for a quite a detailed level of Majorana spectroscopy, reflecting e.g.~their bandwidth, their interactions with the flux pair from zero- all the way to many-particle signals, their arrangement in the perturbed ground state or the presence of bound states. We believe that this is an important feature in the broader quest for Majorana physics, which has been central to topological condensed matter physics for a while. It will remain to be seen to what extent some of the finer features, such as the higher multi-particle continua with their relatively small weight, will in practice be visible. In the first instance, both the  broad and sharp  features at relatively low energies are going to be the most likely signatures, and our exact solution will hopefully be of use in modelling these in an attempt to fit experimental results~\cite{Arnab2015}.

The technology developed here to study the dynamical structure factor may be applied to other members of the large and growing family of Kitaev QSLs~\cite{Yao2009, Hermanns2014, Hermanns2015, Subhro2014, Qi2014}. Majorana spectroscopy as outlined above will likely be an appropriate framework for interpreting the results in most cases, where it is an entirely open question how `details' -- such as spatial dimensionality, dimensionality of the zero-energy Majorana manifold, symmetries and possibly different selection rules --  will manifest themselves. Beyond building up a compendium of possible behaviours, we hope that such a programme will lead to a practically useful field guide for the identification of such QSLs. Considering the properties of surface states of 3D QSLs appears to be another promising line of future work.

Finally, the general methodological advances of our work should be applicable in contexts beyond the KHM. An approximate approach to the computation of the dynamic structure factor, here called the adiabatic approximation, has been known and used elsewhere for a long time, even in the remarkably similar problem of a missing core electron in a 'single graphite plane' \cite{ritskomele}.

Beyond this, however, our calculation of the dynamic structure factor can be taken as an exact solution of a local quantum quench~\citep{Baskaran2007}, related but not identical, to the X-ray edge problem~\citep{Nozieres1969}, and as such represents a contribution to non-equilibrium quantum dynamics in its own right, for which we have developed two complementary approaches.  First, in order to obtain exact results in the thermodynamic limit we have adapted a method from the theory of singular integral equations~\cite{mush,greb,Knolle2014}, which can be extended to calculations of the full frequency dependence for the local impurity quenches with non-standard fermionic density of states. Second, we have derived an exact determinant expression for an arbitrary quadratic quench (allowing for the presence of anomalous terms), which can be useful for numerical calculations with systems considerably larger than those accessible to exact diagonalization. 

We hope that the insights presented in our work, together with the technology developed here, will be of use for the investigation of an extended class of fermionic many-problems.

\section{Acknowledgements}
We thank G.~Baskaran and F.~H.~L.~Essler for helpful discussions. The work of J.K.~is supported by a Fellowship within the Postdoc-Program of the German Academic Exchange Service (DAAD). J.T.C.~is supported in part by EPSRC Grant No.~EP/I032487/1, D.K.~is supported by EPSRC Grant No.~EP/M007928/1. This collaboration was supported by the Helmholtz Virtual Institute ``New States of Matter and their Excitations''.

\appendix
%%%%%%%%%%%%%%%%%%%%%%%%%%%%%%%%%%%%%%%%%%
%%%%%%%%%%%%%%%%%%%%%%%%%%%%%%%%%%%%%%%%%%%
\section{Determinant approach for correlation functions}\label{ExactPfaff}
Below we present the derivation of exact expressions for the dynamical spin correlators, similar in spirit to those of full-counting statistics (FCS). These results can be applied to evaluate numerically the structure factor in finite systems, whose size is much larger than is amenable to exact diagonalization approaches. The results of this Section are not constrained to a specific model, and may provide a starting point for further analytical considerations.
\subsection{Definitions}
The dynamical spin correlation functions of the (extended) KHM can be expressed in terms of the matrix elements, see Eq.~(\ref{SpinCorrFctPfaffian}), which have the following form
\begin{eqnarray}
\label{MatixElementShort}
M_{ql}(t)=\langle M_0|\hat{a}_{q} e^{-i \hat{H}_z t} \hat{a}_{ l}^{\dagger}|M_0\rangle,
\end{eqnarray}
where $\hat{H}_z=\hat{H}_0+\hat{V}_z$ is the Majorana (matter) Hamiltonian in the presence of two flipped fluxes (playing a role of a local potential for Majoranas),
and $|M_0\rangle$ is the ground state of $\hat{H}_0$, defined as $\hat{a}_q|M_0\rangle=0$ for all $q\in \mathrm{BZ}$.

We will now derive the expressions for the matrix elements (\ref{MatixElementShort}), which are suitable for numerical evaluation, in terms of Pfaffians. First, it is convenient to express the Hamiltonian $\hat{H}_z$ in terms of operators $\hat{a}_q$. In this representation the Hamiltonian assumes the Bogoliubov-de Gennes form
\begin{eqnarray}
\label{BogDeGenHamiltonain}
\hat{H}_z=\sum_{ij} \left[ h_{ij}\hat{a}_{i}^{\dagger}\hat{a}_{j} + \frac{1}{2}\Delta^{\dagger}_{ij} \hat{a}_{i}\hat{a}_{j} + \frac{1}{2}\Delta_{ij} \hat{a}_{i}^{\dagger}\hat{a}_{j}^{\dagger}\right],
\end{eqnarray}
where $h,\Delta$ are the matrices to be defined later (the following discussion does not rely on a specific form of the matrices, provided that the Hamiltonian is Hermitian).

The matrix elements defined in Eq.~(\ref{MatixElementShort}) can be calculated using Grassmann path integrals. First, we represent the state $|M_0\rangle$ in terms of a trace over complete set of states $\{|n\rangle\}$ using a projector to the ground state, which has no fermions, so that
\begin{multline}
\label{Matrix1}
M_{ql}(t)=\sum_n \langle n|\hat{a}^{\dagger}_N \hat{a}^{\dagger}_{N-1}\ldots \hat{a}^{\dagger}_1\ \hat{a}_{q} e^{-i\hat{H}_z t} \hat{a}_{l}^{\dagger}\ \hat{a}_1\ldots\hat{a}_N|n\rangle \\ 
= \mathrm{Tr}\{  e^{-i\hat{H}_z t}\hat{a}_{l}^{\dagger}\hat{a}_1\ldots\hat{a}_N\hat{a}^{\dagger}_N\ldots \hat{a}^{\dagger}_1\hat{a}_{q} \}.
\end{multline}
Here $N$ is the total number of momentum states in the Brillouin zone (which is equal to the number of unit cells). After commuting all $\hat{a}_l^{\dagger}$ to the right and all $\hat{a}_q$ to the left, we obtain
\begin{multline}
\label{Matrix3}
M_{ql}(t) = (-1)^{q+l} \mathrm{Tr}\{\hat{a}^{\dagger}_N\ldots\hat{a}^{\dagger}_{l+1}\hat{a}^{\dagger}_{l-1}\ldots\hat{a}^{\dagger}_1\\ \times e^{-i\hat{H}_z t} \hat{a}_1\ldots\hat{a}_{q-1} \hat{a}_{q+1}\ldots\hat{a}_N \} \\ - \delta_{ql}  \mathrm{Tr} \{\hat{a}^{\dagger}_N\ldots\hat{a}^{\dagger}_1 e^{-i\hat{H}_z t} \hat{a}_1\ldots \hat{a}_N \}.
\end{multline}
Notice that a single creation operator $\hat{a}^{\dagger}_l$, and a single annihilation operator $\hat{a}_q$ is absent in the first term. Now our task is to derive a generating functional $\mathbf{F}[J^{*},J]$. Then the matrix elements $M_{ql}$ (and other Green functions) are obtained by a differentiation with respect to the source terms $J$. For example
\begin{equation}
\mathrm{Tr} \{\hat{a}^{\dagger}_N\ldots\hat{a}^{\dagger}_1 e^{-i\hat{H}_z t} \hat{a}_1\ldots\hat{a}_N \}=  \frac{\partial^{2N}\mathbf{F} [J^{*},J]}{\partial J_N ... \partial J_1 \partial J_1^{*}...\partial J_N^{*}}\nonumber.
\end{equation}

Using the identity operator for the set of Grassmann variables $\{\phi_\alpha\},\ \alpha=1\ldots N$ (see e.g.~Ref.~\cite{Negele2008})
\begin{equation}
\hat{I}=\int \prod_{\alpha=1}^N d\phi_\alpha^*d \phi_\alpha e^{-\sum_\alpha \phi_\alpha^* \phi_\alpha} |\phi\rangle\langle \phi|,
\end{equation}
one can write for $\mathcal D_0\equiv\mathrm{Tr}[e^{-i\hat{H}_z t}]$,
\begin{equation}\nonumber
\label{PartitionFct}
\mathcal D_0= \int \prod_{\alpha=1}^N d\phi_\alpha^*d \phi_\alpha e^{-\sum_\alpha \phi_\alpha^* \phi_\alpha} \langle -\phi|e^{-i\hat{H}_z t} |\phi\rangle.
\end{equation}
Using Suzuki-Trotter decomposition on the r.h.s.~of this equation by inserting $M-1$ times the identity operator, so that we have $M$ time intervals ($\delta t= t/M$), we obtain
\begin{multline}
\label{PathInt1}
\mathcal D_0=\lim_{M\to\infty}\int D[\phi^{*},\phi]
e^{-\sum_{l=1}^M \langle\phi^{l}|(\phi^{l}-\phi^{l-1})\rangle}\\
\times e^{-i\delta t\sum_{l=1}^M  [\langle\phi^{l}|\hat{h}|\phi^{l-1}\rangle
 + \frac{1}{2}\langle\phi^{l}|\hat{\Delta}|\phi^{l*}\rangle+\frac{1}{2}\langle\phi^{(l-1)*}|\hat{\Delta}^{\dagger}|\phi^{l-1}\rangle]}.\nonumber
\end{multline}
Here $\int D[\phi^*,\phi]\equiv\int\prod_{l=1}^{M}\prod_{\alpha=1}^N d\phi_\alpha^{l*}d\phi_\alpha^{l}$, and the fields obey anti-periodic boundary conditions $\phi^{M}_{\alpha}=-\phi_{\alpha}^{0}$. Next, we introduce a Fourier transform in the Matsubara space
\begin{equation}
\label{FourierMatsubara}
\phi_\alpha^l = \frac{1}{\sqrt{M}}\sum_{p=-\frac{M}{2}}^{p=\frac{M}{2}-1} {\tilde{\phi}}_\alpha^p e^{i\omega_p l},
\end{equation}
with the Matsubara frequencies given by $\omega_p=\frac{2\pi}{M}(p+\frac{1}{2})$.
The sum over frequencies in the Eq.~(\ref{FourierMatsubara}) can be separated into positive and negative components (note that $\omega_{-(p+1)}=-\omega_{p}$). We absorb the common factor $i \delta t$ into a redefinition of matrices $\tilde{h}=i h \delta t$ and $\tilde{\Delta}=i \Delta \delta t $, and take the large $M$ limit. Note that the latter introduces a phase ambiguity in the path integral, see e.g.~Ref.~\cite{BDS}, which will be resolved at a later stage. By introducing a vector notation (in the index $\alpha$), we combine the Grassmann variables into a single vector $|\Phi^p\rangle=[|\tilde{\phi}^{p} \rangle\ \ |\tilde{\phi}^{-(p+1)*} \rangle]^{T}$, and after defining the matrix
\begin{equation}
\hat{\mathcal{F}}_p=
\begin{bmatrix}
i\omega_p+\frac{1}{2}\tilde{h} & \frac{1}{2}\tilde{\Delta}\\
 \frac{1}{2}\tilde{\Delta}^{\dagger}  & i\omega_p-\frac{1}{2}\tilde{h}^T
\end{bmatrix},
\end{equation}
the expression for the $\mathcal{D}_0$ assumes a compact form
\begin{equation}
\label{PathInt3}
\mathcal D_0 = \lim_{M\to\infty}\int D[\Phi^{\dagger},\Phi]e^{-\sum_{p=0}^{\frac{M}{2}-1}\langle\Phi^p|\hat{\mathcal{F}}_p|\Phi^p\rangle}.
\end{equation}
In order to evaluate the path integral in Eq.~(\ref{PathInt3}), we first diagonalize the quadratic Hamiltonian
\begin{eqnarray}
\label{Bog11}
\hat{\mathcal{H}}  = \frac{1}{2} 
\begin{bmatrix}
\hat{a}_i^{\dagger} &  \hat{a}_i
\end{bmatrix} 
\begin{bmatrix}
\tilde h & \tilde \Delta\\
 \tilde \Delta^{\dagger}& - \tilde h^T
\end{bmatrix}
\begin{bmatrix}
\hat{a}_j\\
 \hat{a}_j^{\dagger}
\end{bmatrix} \equiv
\hat \alpha^{\dagger} \bar H \hat \alpha    
\end{eqnarray}
using a Bogoliubov transformation described by a matrix $\hat T$, see Ref.~\citep{ripka}, such that
\begin{eqnarray}
\label{Bog2}
\hat T \bar H \hat T^{-1} = \frac{1}{2}
\begin{bmatrix}
\hat \Omega & 0\\
 0   & - \hat \Omega
\end{bmatrix}%\hat \beta = \hat T \hat \alpha \ \ \text{and} \ \ 
\end{eqnarray}
where $\hat\Omega$ is a $N\times N$ diagonal matrix of eigenvalues $\Omega_n$. Note that the eigenvalues $E_n$ in Eq.~(\ref{UnitaryTrafo}) and the eigenvalues $\Omega_n$ in Eq.~(\ref{Bog2}) differ by a factor of $i \delta t$. We can now write the expression for $\hat{\mathcal{H}}$ in the diagonal form
$\hat{{\mathcal H}}  =  \sum_{n>0} \Omega_n \hat{b}_n^{\dagger}\hat{b}_n  - \frac{1}{2} \sum_{n>0}  \Omega_n$. In the remainder we will omit the constant contribution, which  will be restored in the final expression.
The matrices of the Bogoliubov transformations have been defined in Sec.~IV, and have a form
\begin{equation}
\label{Tmatrix}
\hat T^{-1} =  
\begin{bmatrix}
\mathcal{X}^T & \mathcal{Y}^{\dagger}\\
 \mathcal{Y}^T   & \mathcal{X}^{\dagger}
\end{bmatrix}
\ \ \text{and}  \ \ 
\hat T =  
\begin{bmatrix}
\mathcal{X}^* & \mathcal{Y}^*\\
\mathcal{Y}   & \mathcal{X}
\end{bmatrix}
\end{equation}
with $\mathcal{X},\mathcal{Y}$ being $N\times N$ matrices. We recall that this transformation relates the operators $\hat{a}_\alpha$ and $\hat{b}_\alpha$, which diagonalize the Hamiltonian $\hat{H}_0$ in the flux-free sector, and the Hamiltonian $\hat{H}_z$ in the two-flux sector respectively. Now we can write Eq.~(\ref{PathInt3}) in a diagonal form. After introducing another vector combined of Grassman variables $\langle \Psi^p|\equiv[\langle \psi_{1}^{p*}|\ \langle \psi_{2}^p|]=\langle \Phi^p| \hat T^{-1}$ we obtain
\begin{multline}
\label{PathInt4}
 {\mathcal D_0} = 
 \int D[\Psi^{\dagger},\Psi]  
 e^{- \sum_{p=0}^{\infty} 
\langle \Psi^{p}| 
\begin{bmatrix}
 i\omega_p+\hat \Omega & 0 \\
 0 & i\omega_p-\hat \Omega
\end{bmatrix}
|\Psi^p\rangle} \\ 
=\prod_{n=1}^{N}e^{\sum_{p=-\infty}^{\infty} \ln{(i\omega_p+\Omega_n)}} = 
\prod_{n=1}^{N}(1+e^{-i E^F_n t}).
\end{multline}
The last line was obtained by evaluating the Grassmann integrals, see e.g.~\cite{BDS}, and we used a standard formula $\sum_p \ln{(i\omega_p+\Omega_n)}=\ln[1+e^{-i E_n t}]$ (note that $\Omega_n=i \delta t E_n^F$).

\subsection{Generating functional}
We construct a generating functional by adding source terms, represented by vectors of Grassman variables $|\tilde{\mathcal{J}}^p\rangle= [ |\tilde{J}^{p} \rangle, - |\tilde{J}^{-(p+1)*}\rangle]^{T}$,  to the path integral
\begin{multline}
\label{GeneratingFunctional1}
\mathbf{F} [J^{*},J]  =   \int D[\Phi^{\dagger},\Phi] e^{- \sum_p 
\langle\Phi^p|\hat{\mathcal{F}}_p|\Phi^p\rangle}\\
\times e^{\sum_{p}\langle\Phi^p|\tilde{\mathcal{J}}^p\rangle + \langle \tilde{\mathcal{J}}^p| \Phi^p\rangle}.
\end{multline}
The Gaussian integrals are calculated using Bogoliubov transformation, and
after taking the inverse Fourier transform back from the Matsubara frequency space
\begin{equation}
\tilde{J}_\alpha^{p}  =  \frac{1}{\sqrt{M}}\sum_{m=1}^M e^{i\omega_p m} J_\alpha^{m},
\end{equation}
we arrive at the expression
\begin{equation}\nonumber
%\label{genfunct3}
 \mathbf{F} [J^{*},J]= \mathcal{D}_0  e^{
 \sum\limits_{p,mn} \langle \mathcal{J}^{m}| \hat T^{\dagger} 
\begin{bmatrix}
\frac{e^{-i\omega_p (m-n)}}{i\omega_p+\hat{\Omega} }& 0\\
0  & \frac{e^{+i\omega_p (m-n)}}{i\omega_p-\hat{\Omega}}
\end{bmatrix}
\hat T | \mathcal{J}^{n}\rangle}. 
\end{equation}
In the following we will only require the functional derivatives taken at the times having the index $M$, e.g.~the matrix elements of the form
$\partial^{2}{\mathbf  F}/\partial J_i^M \partial J_j^{M*}$,
because all operators $\hat{a}_i^{\dagger},\hat{a}_j$ in 
$ \mathrm{Tr} \{ \ldots \hat{a}_i^{\dagger} \ldots  e^{-i\hat{H}_z t} \ldots \hat{a}_j \ldots \}$  are taken at the same time (at the boundary, see Eq.~(\ref{PartitionFct})), and we set $m=n=M$. 
Further we extend the sums over $p$ negative frequencies, and introduce the definitions
$\sum_{p=-\infty}^{\infty}[i\omega_p \pm \hat{\Omega}]^{-1} \equiv n^{\mp}(\hat{E})$, where $\hat E$ is s vector of eigenvalues $E_n$.
The phase ambiguity in the path integral can be fixed by comparing the results with the  ones calculated by standard operator theory of low-order matrix elements, e.g.~$\mathrm{Tr}\{ \hat{a}_l^{\dagger} e^{-i\hat{H}_z t}\hat{a}_q\}$, see Ref.~\citep{Knolle2014c}. The functions
\begin{equation}
n^{\mp} (\hat{E})=\frac{1}{1+e^{\pm \hat{E}}}
\end{equation}
assume the form of the Fermi-distribution, and we obtain
\begin{equation}
{\mathbf  F} [J^{*},J]  = 
 {\mathcal D}_0  e^{\frac{1}{2}  
\langle \mathcal{J}|
\hat T^{-1} 
\begin{bmatrix}
n^-(\hat{E})& 0\\
0  & n^+(\hat{E})
\end{bmatrix}
\hat T |\mathcal{J}\rangle}.
\end{equation}
Note the factor of $1/2$, which is due to the fact that the sums have been extended to negative frequencies; we have dropped the index $M$. The matrix can now be written in explicit form
\begin{eqnarray}
\label{MatrixGenFct}
& \hat T^{-1} 
\begin{bmatrix}
n^-(\hat{E})& 0\\
0  & n^+(\hat{E})
\end{bmatrix}
\hat T  = 
\begin{bmatrix}
\mathcal{A}& \mathcal{B}\\
\mathcal{C}  & \mathcal{D}
\end{bmatrix},\nonumber
\end{eqnarray}
where the entries $\mathcal{A,B,C,D}$ are given by $N\times N$ matrices
\begin{eqnarray}
\mathcal{A}  &=&  \mathcal{X}^T n^- \mathcal{X}^* +\mathcal{Y}^{\dagger} n^+ \mathcal{Y},\ \ \
\mathcal{B}  =  \mathcal{X}^T n^- \mathcal{Y}^* +\mathcal{Y}^{\dagger} n^+ \mathcal{X},\nonumber \\ 
\mathcal{C}  &=&  \mathcal{Y}^T n^- \mathcal{X}^* +\mathcal{X}^{\dagger} n^+ \mathcal{Y},\ \ \
\mathcal{D}  =  \mathcal{Y}^T n^- \mathcal{Y}^* +\mathcal{X}^{\dagger} n^+ \mathcal{X}. \nonumber
\end{eqnarray}
In terms of these matrices, the partition functions reads
\begin{equation}
\label{part_fun}
\mathbf {F} [J^{*},J]  = 
 {\mathcal D_0}e^{\frac{1}{2}\left\{ 
\langle J|\hat{\mathcal{A}}|J\rangle -\langle J| \hat{\mathcal{B}} J^{*}\rangle - \langle J^{*}|\hat{\mathcal{C}}|J\rangle +\langle J^{*}|\hat{\mathcal{D}}|J^{*}\rangle\right\} }
\end{equation}
where we substituted  $\mathcal{D}$ with $-\mathcal{A}^{T}$, which fixes the phase ambiguity. Note that in the absence of the anomalous terms in the Hamiltonian, the expression for the generating functional that we obtained reduces to the standard free-fermion result~\citep{Negele2008}.

One could now in principle calculate the matrix elements in Eq.~(\ref{Matrix3}) by differentiating Eq.~(\ref{part_fun}) with respect to the sources. However we find it more convenient to proceed in a different way. First, we reorder the expression under the exponent into a trace with the anti-symmetric matrix $\mathcal S=-\mathcal S^{T}$, defined as
\begin{equation}
\mathcal S\equiv\begin{bmatrix}
\mathcal{-B} & \mathcal{A}\\
-\mathcal{A}^T & \mathcal{-C}
\end{bmatrix},
\end{equation}
so that we can write
\begin{align}
\label{GeneratingFunctionalAntiSymmetric}
 {\mathbf  F} [J^{*},J]  = 
 {\mathcal D}_0 e^{\frac{1}{2}\langle \mathcal{J}|\mathcal S|\mathcal{J}\rangle}.
\end{align}
 
\subsection{Matrix Elements}
Let us now discuss the calculation of the matrix element
\begin{multline}
L(t)\equiv\mathrm{Tr} \{\hat{a}^{\dagger}_N\ldots\hat{a}^{\dagger}_1 e^{-i\hat{H}_z t} \hat{a}_1\ldots\hat{a}_N\} \nonumber  \\ 
=(-1)^N \int D[\phi^{*},\phi] e^{-S[\phi*,\phi]} \phi_N^{M*} \ldots \phi_1^{M*}\phi_1^{0}\ldots\phi_N^{0} \nonumber\\   
=  \frac{\partial^{2N}}{\partial J_N ... \partial J_1 \partial J_1^{*}...\partial J_N^{*}} \mathbf {F} [J^{*},J],
\end{multline}
where the sign factor $(-1)^{N}$ on the second line appears due to anti-periodic boundary conditions on Grassmann variables.
Differentiation with respect to Grassmann variables is essentially the same as integration (see Ref.~\citep{Zinn2002}) so that\begin{equation}
L(t)  = \int dJ_N \ldots dJ_1 dJ_1^* ...dJ_N^*\ {\mathbf  F} [J^{*},J].
\end{equation}
After relabelling $[J_1^*\ldots J_N^*J_1 \ldots J_N] \rightarrow [\theta_1 ... \theta_{2N}]$ we arrive at the result
\begin{multline}
L(t)=(-1)^\frac{N(N-1)}{2}{\mathcal D}_0
\times\int d\theta_{2N}\ldots d\theta_1 e^{\frac{1}{2}\theta^{T} \mathcal S \theta} 
\\=(-1)^{\frac{N(N-1)}{2}}{\mathcal D}_0\  \mathrm{Pf}\mathcal S.
\end{multline}
In the last line we used Gaussian integration of anti-symmetric matrices with Grassmann variables, which results in a Pfaffian, see e.g.~Ref.~\citep{Zinn2002}.

The first term in the matrix element $M_{ql}$ in Eq.~(\ref{Matrix3}) has the form with a creation and an annihilation operator removed, namely
 \begin{multline}
 R_{ql}(t)\equiv\mathrm{Tr} \{\hat{a}^{\dagger}_N\ldots\hat{a}^{\dagger}_{l+1}\hat{a}^{\dagger}_{l-1}\ldots\hat{a}^{\dagger}_1 \\ \times e^{-i\hat{H}_z t} \hat{a}_1\ldots\hat{a}_{q-1}\hat{a}_{q+1}\ldots\hat{a}_N \} \nonumber \\= (-1)^{\frac{(N-1)(N-2)}{2}}{\mathcal D}_0 \int D[\theta_{ql}] e^{\frac{1}{2}\theta^{T} \mathcal S\theta},
 \end{multline}
where the measure of the integral is defined as 
\begin{equation}\nonumber
D[\theta_{ql}]=d\theta_{2N}\ldots d\theta_{2N-l+1}d\theta_{2N-l-1}\ldots d\theta_{q+1}d\theta_{q-1}\ldots d\theta_{1}.
\end{equation}
In the series expansion of the exponent, the only non-vanishing terms which remain after the integration are those which contain $2N-2$ Grassmann operators
\begin{multline}
 \int D[\theta_{ql}] e^{\frac{1}{2}\theta^{T} \mathcal{S}\theta} \nonumber
 =\frac{2^{-N+1}}{(N-1)!} \int D[\theta](\theta^{T}\mathcal{S}\theta)^{N-1}\nonumber\\ 
 =  \frac{2^{-N+1}}{(N-1)!} \sum_{P}
\mathrm{sgn}[P]\mathcal{S}_{i_1 i_2}\mathcal{S}_{i_3 i_4}\ldots
 = \mathrm{Pf}\mathcal {S}_{[ql]},
\end{multline}
where $P$ is a transposition of $2N-2$ indices (i.e. $1\ldots2N$ excluding $q$ and $2N-l$. The matrix $\mathcal{S}_{[ql]}$ is obtained from the matrix $\mathcal{S}$ by removing two lines and two columns at positions $2N-l$ and $q$. Finally we arrive at the exact expression for the matrix element
\begin{equation}
\label{MatrixElementOpRemoved}
R_{ql}(t)=(-1)^{\frac{(N-1)(N-2)}{2}}{\mathcal D}_0\ \mathrm{Pf}\mathcal{S}_{[ql]}.
\end{equation}

The Eq.~(\ref{MatrixElementOpRemoved}) requires evaluation of a Pfaffian of the matrix $\mathcal{S}_{[ql]}$ with size $2(N-1)\times 2(N-1)$. 
If numerical implementation of this equation the determinant $|{\mathcal D_0}|$ can become very large, while the absolute value of Pfaffians very small, which would bring large numerical errors. In order to regularise the matrix elements one can use the well-known property of Pfaffians, namely $\mathrm{Pf}[B A B^T]= \mathrm{Det}B\times \mathrm{Pf}A$ for any matrix $B$ of the same size as $A$. To this end we introduce a $2N\times2N$ diagonal matrix $B=\mathrm{diag}[(1+e^{-it E^F_n}),\underbrace{1\ldots1}_{N \mathrm{times}}]$, whose determinant is equal to
\begin{equation}
\label{DeterminantNeu}
\mathrm{Det}B=\prod_n (1+e^{-it E^F_n})={\mathcal D}_0.
\end{equation}
This matrix provides a required regularisation of the matrix elements through the expression
\begin{equation}
\mathcal{D}_0\mathrm{Pf}\mathcal S = \mathrm{Pf}[B \mathcal S B].
\end{equation}
While the matrix under the Pfaffian on the l.h.s.~of this expression can become ill-conditioned, the r.h.s.~of this expression can be evaluated numerically without difficulties.

For the Pfaffian entering expression for the $R_{ql}(t)$, we use the expansion formula of Eq.~(\ref{PfaffianDet1}) to derive the identity 
\begin{equation}
 \label{PfaffianProperty1}
\mathrm{Pf}\mathcal{S}_{[ql]}= 
(-1)^{l+q}(-1)^{N+1} 
\mathrm{Pf}
\underbrace{\begin{bmatrix}
 & 0 &  & 0 &  \\
0 & 0& 0 &+1 & 0\\
 & 0 & & 0& \\
0 &-1 &0 &0 & 0\\
 &0 & &0 &
\end{bmatrix}}_{\equiv\mathcal{S}_{\{ql\}}},
\end{equation}
where the $2N\times2N$ matrix $\mathcal{S}_{\{ql\}}$ on the r.h.s.~is obtained from the matrix $\mathcal{S}$ by setting the rows and columns $2N-l$ and $q$ to zero, and  $\mathcal S_{q\ 2N-l}=-1$ and $\mathcal S_{2N-l\ q}=+1$. We have
\begin{equation}
\label{MatrixElementOpReplaced0}
R_{ql}(t)  = (-1)^{q+l}
 (-1)^{\frac{N(N-1)}{2}}{\mathcal D_0} \mathrm{Pf}[ \mathcal{S}_{\{ql\}}],
\end{equation}
where the Pfaffian is calculated for a $2N\times 2N$ matrix. 

As was shown above for the matrix $\mathcal{S}$, one can regularise the Pfaffian $\mathrm{Pf}\mathcal{S}_{\{ql\}}$ using the same matrix $B$. After defining a phase-factor
\begin{equation}
K(t)=e^{-i E_{0}^F t} (-1)^{\frac{N(N-1)}{2}},
\end{equation}
and collecting all contributions, we obtain the expression for the complete matrix element  
\begin{equation}\label{pfaffian1}
M_{ql}(t) = K(t)\{ \mathrm{Pf}[B \mathcal{S}_{\{ql\}}B] - \delta_{ql}  \mathrm{Pf}[B \mathcal{S} B]\},
\end{equation}
which is Eq.~(\ref{MatrixElementFinal}) of the main text. This expression can be further simplified by generalising a theorem for Pfaffians, see Ref.~\cite{kanz}, and we obtain
\begin{equation}\label{pfaffian2}
M_{ql}(t)=-K(t)\mathcal{D}_0\mathrm{Pf}\mathcal{S}\times[\hat{I}_{N}+\hat{G}^{T}]_{ql},
\end{equation}
where $\hat{G}$ is the upper right $N\times N$ block of the inverse of $\mathcal S$, and $I_N$ is the $N\times N$ identity matrix. One can still reduce the size of the matrices which needs to be inverted by a factor of two due to a special structure of the result. Let us introduce a $N\times N$ matrix
\begin{equation}
\Lambda=\mathcal{Y}^T_Fe^{-i \hat{E}^F t}\mathcal{Y}_F^{*}+\mathcal{X}_F^{\dagger}e^{i \hat{E}^{F} t}\mathcal{X}_F,
\end{equation}
where $\hat{E}^{F}$ is a $N\times N$ diagonal matrix of $E^{F}_n$. We arrive at the following simple result
\begin{equation}\label{detexact}
M_{ql}(t)=\sqrt{\mathrm{Det}[\Lambda(t)]}[\Lambda^{-1}(t)]_{ql},
\end{equation}
where a precise definition of the square root is the following
\begin{equation}
\sqrt{\mathrm{Det}[\Lambda(t)]}=\sqrt{|\mathrm{Det}[\Lambda(t)]|}e^{i \varphi_\Lambda(t)/2},
\end{equation}
and the phase $\varphi_\Lambda(t)=\arg[\mathrm{Det}[\Lambda(t)]]$ is taken to be a continuous function of time. Note also that $M_{ql}(0)=\delta_{ql}$. The phase $\varphi_\Lambda(t)$ contains a large linear part $-2E_0 t$, and in numerical calculations it is convenient to separate this contribution first. In fact the loop contribution in the diagrammatic expansion can be written as
\begin{equation}\label{loops}
\langle\hat{\mathcal{S}}(t,0)\rangle=e^{i E_0 t}\langle e^{-i\hat{H}_z t}\rangle=e^{i E_0 t}\sqrt{\mathrm{Det}[\Lambda(t)]}.
\end{equation}

\section{Some useful properties of Pfaffians}\label{AppPfaffianIntro}
The results of the previous Section have been expressed in terms of Pfaffians, which often appear in Gaussian integrals over anti-commuting variables, see e.g.~\citep{Zinn2002}. Below we present a short overview of the definitions and the properties of Pfaffians that are relevant for our calculations. 

A Pfaffian is an extension of a determinant for skew-symmetric matrices $A=-A^{T}$. It is always possible to write a determinant of a skew-symmetric matrix as the square of a polynomial in the matrix elements, \citep{Zinn2002}
\begin{eqnarray}
\label{PfaffianDet}
[\mathrm{Pf}A]^2 =\mathrm{Det}A. 
\end{eqnarray}
The formal definition of a Pfaffian for a $2N \times 2N$ skew-symmetric matrix $A$ is
\begin{equation}\nonumber
\label{PfaffianDetFomral}
\mathrm{Pf}A = \frac{1}{2^N N!} \sum_{P\in{i_1\ldots i_{2N}}} \mathrm{sgn}[P]\ a_{i_1 i_2} a_{i_3 i_4} \ldots a_{i_{2N-1} i_{2N}}  
\end{equation}
with the matrix elements $a_{ij}$, and $\mathrm{sgn}[P]=\pm 1$ the sign of the permutation $P$. 
Hence, the Paffian has the unique sign for the square root $\mathrm{Pf}A=\pm\sqrt{\mathrm{Det}A}$.
Note that the Pfaffian of an odd-dimensional matrix is zero.

Several properties known from determinants carry over in a modified way to Pfaffians \citep{Zinn2002,Wimmer2011}:
\begin{itemize}
 \item 
Multiplication of a row and a column on a constant is the same as multiplication of the entire Pfaffian on this constant,
 \item   
Interchanging two rows and the corresponding columns flips the sign of the Pfaffian,
 \item
Adding  multiples of a row and the corresponding column to another row and a column does not affect a Pfaffian.
\end{itemize}
Another very useful property (in particular for numerical computations) is an expansion formula for the Pfaffians
\begin{eqnarray}
\label{PfaffianDet1}
\mathrm{Pf}A = \sum_{i=2}^{2N} (-1)^i a_{1i}\ \mathrm{Pf}[A_{1i}], 
\end{eqnarray}
with the reduced matrix $A_{1i}$ having the first row and the $i$-th column removed. In addition, we will use the relation
\begin{equation}
\label{PfaffianProp2}
\mathrm{Pf}[B A B^T] = \mathrm{Pf}A\ \mathrm{Det}B,
\end{equation}
which is valid for an arbitrary square matrix $B$ having the same dimension as $A$.

\section{Ground state overlap from the Bogoliubov transformations}\label{FluxToFlux}
The Bogoliubov transformations for the matter fermion operators in a given flux sector (in a fixed gauge) read \cite{ripka} 
\begin{eqnarray}
\label{BogoliubovInComponents}
\hat f_i & = X^T_{ik} \hat a_k + Y^{\dagger}_{ik} \hat a_k^{\dagger}, \\ \nonumber
\hat f_j^{\dagger} & = Y^T_{jl} \hat a_l + X^{\dagger}_{jl} \hat a_l^{\dagger},
\end{eqnarray}
where $X$,$Y$ are $N\times N$ matrices, and we assumed summation over repeating indices. 
The calculation of the dynamical structure factor requires matrix elements between states in different flux sectors. Let $\hat b$ and $\hat a$ be the operators in inequivalent flux sectors in which Hamiltonians for the matter fermions is diagonal. For definiteness let's take a system with (index $F$) and without (index $0$) fluxes. The Hamiltonian in each case is diagonalised by respective Bogoliubov transformations, namely in the flux-free sector we have
\begin{eqnarray}
\label{BogNoFlux}
\begin{pmatrix}
X_0^* & Y_0^*\\
Y_0 & X_0
\end{pmatrix}
\begin{pmatrix}
\hat f\\
\hat f^{\dagger}
\end{pmatrix} =
\begin{pmatrix}
\hat a\\
\hat a^{\dagger}
\end{pmatrix} 
\end{eqnarray}
with the inverse
\begin{eqnarray}
\begin{pmatrix}
 X_0^T &  Y_0^{\dagger}\\
 Y_0^T &  X_0^{\dagger}
\end{pmatrix}
\begin{pmatrix}
 \hat a\\
 \hat a^{\dagger}
\end{pmatrix} = 
\begin{pmatrix}
\hat f\\
\hat f^{\dagger}
\end{pmatrix} .
\end{eqnarray}
In the sector with fluxes
\begin{eqnarray}
\label{BogFlux}
\begin{pmatrix}
 X_F^* &  Y_F^*\\
 Y_F &  X_F
\end{pmatrix}
\begin{pmatrix}
\hat f\\
\hat f^{\dagger}
\end{pmatrix} =
\begin{pmatrix}
 \hat b\\
 \hat b^{\dagger}
\end{pmatrix}.
\end{eqnarray}
Now one can use these to express $\hat b$ operators in terms of $\hat a$'s
\begin{eqnarray}
\label{FluxNoFluxRelation}
\begin{pmatrix}
 \mathcal{X}^* & \mathcal{Y}^*\\
\mathcal{Y} & \mathcal{X}
\end{pmatrix}
\begin{pmatrix}
\hat a\\
\hat a^{\dagger}
\end{pmatrix} =
\begin{pmatrix}
 \hat b\\
 \hat b^{\dagger}
\end{pmatrix}.
\end{eqnarray}
Here we introduced the matrices
\begin{eqnarray}
\label{FluxNoFluxRelationDefinition}
\begin{pmatrix}
 \mathcal{X}^* & \mathcal{Y}^*\\
\mathcal{Y} & \mathcal{X}
\end{pmatrix}
=
\begin{pmatrix}
X_F^*  X_0^T+ Y_F^* Y_0^T & X_F^*  Y_0^{\dagger}+ Y_F^*  X_0^{\dagger}\\
Y_F  X_0^T+X_F  Y_0^T & Y_F  Y_0^{\dagger}+X_F  X_0^{\dagger}
\end{pmatrix}.
\end{eqnarray}

The ground states are related by the unitary transformation $T$ of Eq.~(\ref{FluxNoFluxRelationDefinition}), and their relative parity is even or odd if the determinant of real orthogonal matrix 
\begin{eqnarray}
\label{Parity}
\mathcal{B}=U T U^{\dagger} \  \text{and} \  \ U=\sqrt{\frac{i}{2}}
\begin{pmatrix}
1 & -i \\
i & -1
\end{pmatrix}.
\end{eqnarray}
is equal to $\pm1$.
Provided that the ground states of Majoranas in the sectors with and without fluxes have the same parity, the two-flux ground state $|M_{F}\rangle$ defined as $\hat b |M_{F}\rangle =0$, is related to the flux-free ground state $\hat a |M_0\rangle =0$ via the following equation, see Ref.~\citep{ripka}
\begin{eqnarray}
\label{GSrelation}
 |M_F\rangle & = &  \mathrm{Det}\left( \mathcal{X}^{\dagger} \mathcal{X}\right)^{\frac{1}{4}} e^{-\frac{1}{2} \mathcal{F}_{ij} \hat a_i^{\dagger} \hat a_j^{\dagger}} |M_0\rangle  \\ \mathcal{F}_{ij} & = &[\mathcal{X}^{* -1} \mathcal{V}^*]_{ij}.
\end{eqnarray}
Hence the overlap between two ground states reads
\begin{eqnarray}
\label{Overlap}
 \langle M_F|M_0 \rangle & = & \mathrm{Det}\left( \mathcal{X}^{\dagger} \mathcal{X}\right)^{\frac{1}{4}}.
\end{eqnarray}
A related overlap arises in the X-ray edge problem between electron ground states with and without a local potential.

\section{Multiparticle contributions to $S(\mathbf{q},\omega)$}
\subsection{Zero- and two-particle contributions}\label{SectionZeroPart}
In order to calculate the $\delta$-function contribution in case (II) one has to modify slightly Eq.~(\ref{Lehmann}). The previous Appendix \ref{FluxToFlux} explained how to relate different flux sectors via proper Bogoliubov rotations, in particular Eq.~(\ref{GSrelation}) establishes the relation between ground states of different flux sectors. Importantly, the product of the parities of two ground states can be found from Eq.~(\ref{Parity}). It is clear from Eq.~(\ref{GSrelation}) that $|M_F\rangle$ generated using a proper Bogolyubov rotation has the same parity as $|M_0\rangle$, so that for case (II) the approach has to be extended to the situation in which the parity changes, which we do in the following.

The problem with the naive use of the Lehmann representation can be traced back to the fact that we have not projected the identity operator to the physical subspace. In order to cure this problem one can re-introduce projectors, or alternatively use improper Bogoliubov transformations. However, there is a simpler way, which is to take advantage of the gauge structure of the Kitaev model. As we discussed above, the model conserves parity of the total number of fermions $N=N_f+N_{\chi}$. A gauge transformation changes the parity of bond and matter fermions while keeping the total parity intact. We note that only the relative parity of matter fermion ground states in two flux sectors (which differ by local fluxes) is important. Since Majorana fermions are their own adjoints $\hat c^{2}_{A}=1$, a correct form of the Lehmann expansion can be obtained by plugging a modified identity operator $\hat c_{A}\sum_{\tilde\lambda} |\tilde\lambda\rangle \langle \tilde\lambda|\hat c_{A}$ into Eq.~(\ref{SpinCorrFct1}), which gives for case (II)
\begin{equation}
\label{LehmannGaugeChange}
S_{AB}^{zz}(t) = -i e^{i E_0t}\sum_{\tilde\lambda} \langle M_0 | e^{-i\hat{H}_{xy}t}| \tilde\lambda \rangle \langle \tilde\lambda |\hat{c}_{A} \hat{c}_{B} | M_0 \rangle.  
\end{equation}
Here we used the fact that $\hat c_{A}  e^{-i \hat{H}_z t}\hat c_{A}=  e^{-i\hat c_{A}\hat{H}_z\hat c_{A}t}$ with \begin{equation}
\hat{H}_{xy}\equiv\hat c_{A}\hat{H}_z\hat c_{A}=\hat{H}_0+\hat{V}_x+\hat{V}_y
\end{equation}
is a gauge transformation, i.e.~it does not alter the flux sector. The Hamiltonian $\hat{H}_{xy}$ can be obtained from $\hat{H}_z$ by reversing signs of all nearest neighbour, and next-nearest neighbourur hoppings on bonds sharing a site $A$. The eigenstates of $\hat{H}_{xy}$ form a set of many-body eigenstates $|\tilde\lambda\rangle$, which can be generated by creating excitations on top of the ground state with fluxes, which it is convenient to represent in terms of the ground state without fluxes via Eq.~(\ref{GSrelation}). Note that the spectrum $E^F_{\tilde\lambda}$ is invariant under the gauge-transformation discussed above, but the parity of its ground state $|M_F^{xy}\rangle$ is  opposite to the parity of the ground state of $\hat{H}_z$ namely $|M_F\rangle$. Hence, in the case (II), i.e.~with vanishing overlap $\langle M_F|M_0 \rangle$ due to different parities, the overlap of the gauge-transformed ground state~is finite, and is given by
\begin{equation}
|\langle M_F^{xy}|M_0 \rangle|^2 =\mathrm{Det}|\mathcal{X}^{xy}|,
\end{equation}
where the matrix $\mathcal{X}^{xy}$ is defined in the Appendix A.

The zero-particle contribution is obtained by restricting the sum over ${\tilde\lambda}$ in Eq.~(\ref{LehmannGaugeChange}) to the ground state
\begin{multline}
\label{LehmannGaugeChange1}
S_{AB}^{zz(0)}(t)   =  -i e^{it(E_0-E^F_{0})} \langle M_0 |M_F^{xy} \rangle \langle M_F^{xy} | \hat c_{A} \hat{c}_{B} | M_0 \rangle \nonumber \\ 
  = e^{it (E_0-E_0^F)} |\langle M_F^{xy}|M_0 \rangle|^2  \{ 1-2[Y^{\dagger}(Y -\mathcal{F}^* X^*)]_{00} \},
\end{multline}
so that the structure factor at $\mathbf{q}=0$ reads
\begin{multline}
S^{zz(0)}_{\mathbf{q}=0}(\omega)  = 8 \pi  
|\langle M_F^{xy}|M_0 \rangle|^2 \delta(\omega -\Delta)\\  \times  \{ 1-[Y^{\dagger}Y]_{00} - \mathrm{Re}[ Y^{\dagger} \mathcal{F}^*X^*]_{00}\}.
\end{multline}
In order to satisfy the parity constraint, the next non-vanishing contribution to Eq.~(\ref{LehmannGaugeChange}), in addition to the $\delta$-function, arises from two-particle excited states
\begin{multline}\label{LehmannTwoPartfinal}
S^{zz(2)}_{\mathbf{q}=0}(\omega)  =  8\pi |\langle M_F^{xy}|M_0 \rangle|^2\\ \times\sum_{\lambda\lambda'}\delta( \omega -[E_{\lambda}^F+E_{\lambda'}^F+\Delta])  \\ 
\times \{  |G_{\lambda\lambda'}|^4+  \mathrm{Re}[ i G^4_{\lambda\lambda'}G^{2*}_{\lambda\lambda'}]/2\}
\end{multline}
with matrix elements 
\begin{multline}
\label{G2}
 G^{[2]}_{\lambda \lambda'} =\langle M_0|b^{\dagger}_{\lambda} b^{\dagger}_{\lambda'}|M_F^{xy} \rangle  \\=  \langle M_F^{x,y}|M_0 \rangle \left\lbrace   \mathcal{Y}_{\lambda l} \mathcal{X}^T_{l \lambda'} + \mathcal{Y}_{\lambda l} \mathcal{F}_{l k} \mathcal{Y}^T_{k \lambda'}\right\rbrace,
\end{multline}
and
\begin{multline}
G^{[4]}_{\lambda \lambda'} = \langle M_0|\hat{c}_{B0} \hat{c}_{A0} b^{\dagger}_{\lambda} b^{\dagger}_{\lambda'}|M_F^{xy} \rangle  \\= -i G^{[2]}_{\lambda \lambda'} +  g^{[4]}_{\lambda \lambda'} +  \tilde g^{[4]}_{\lambda \lambda'}.
\end{multline}
Here we also defined the following contributions
\begin{multline}
\label{G4}
g^{[4]}_{\lambda \lambda'} = 2 i \langle M_F^{xy}|M_0 \rangle \{  
\mathcal{X}_{\lambda i} X_{i0} \mathcal{X}_{\lambda' j} Y_{j0}\\ - \mathcal{X}_{\lambda i} Y_{i0} \mathcal{X}_{\lambda' j} X_{j0} + Y_{i0}^{*} Y_{0i} \mathcal{Y}_{\lambda l} \mathcal{X}^T_{l \lambda'}  \},
\end{multline}
and
\begin{multline}
\tilde g^{[4]}_{\lambda \lambda'}  =  2 i \langle M_F^{xy}|M_0 \rangle  \{
\mathcal{Y}_{\lambda l} \mathcal{X}_{l \lambda'}^T  X_{0j}^T \mathcal{F}_{ji} Y_{i0} \\+ \mathcal{Y}_{\lambda l} \mathcal{F}_{li} X^T_{0i} \mathcal{X}_{\lambda' j} Y_{j0} + \mathcal{Y}_{\lambda l} \mathcal{F}_{li}^T Y_{i0} \mathcal{X}_{\lambda' j} X_{j0} +\\ 
 \mathcal{X}_{\lambda i} X_{i0} \mathcal{Y}_{\lambda' l} \mathcal{F}_{lj} Y_{j0} - 
 \mathcal{X}_{\lambda i} Y_{i0} \mathcal{Y}_{\lambda' l} \mathcal{F}_{lj} X_{j0}\\ -
\mathcal{Y}_{\lambda l} \mathcal{F}_{lk} \mathcal{Y}_{k \lambda'}^T Y_{0j}^T Y_{j0}^* \}.
\end{multline}

\begin{figure}[tb]
\begin{centering}
\includegraphics[width=\columnwidth]{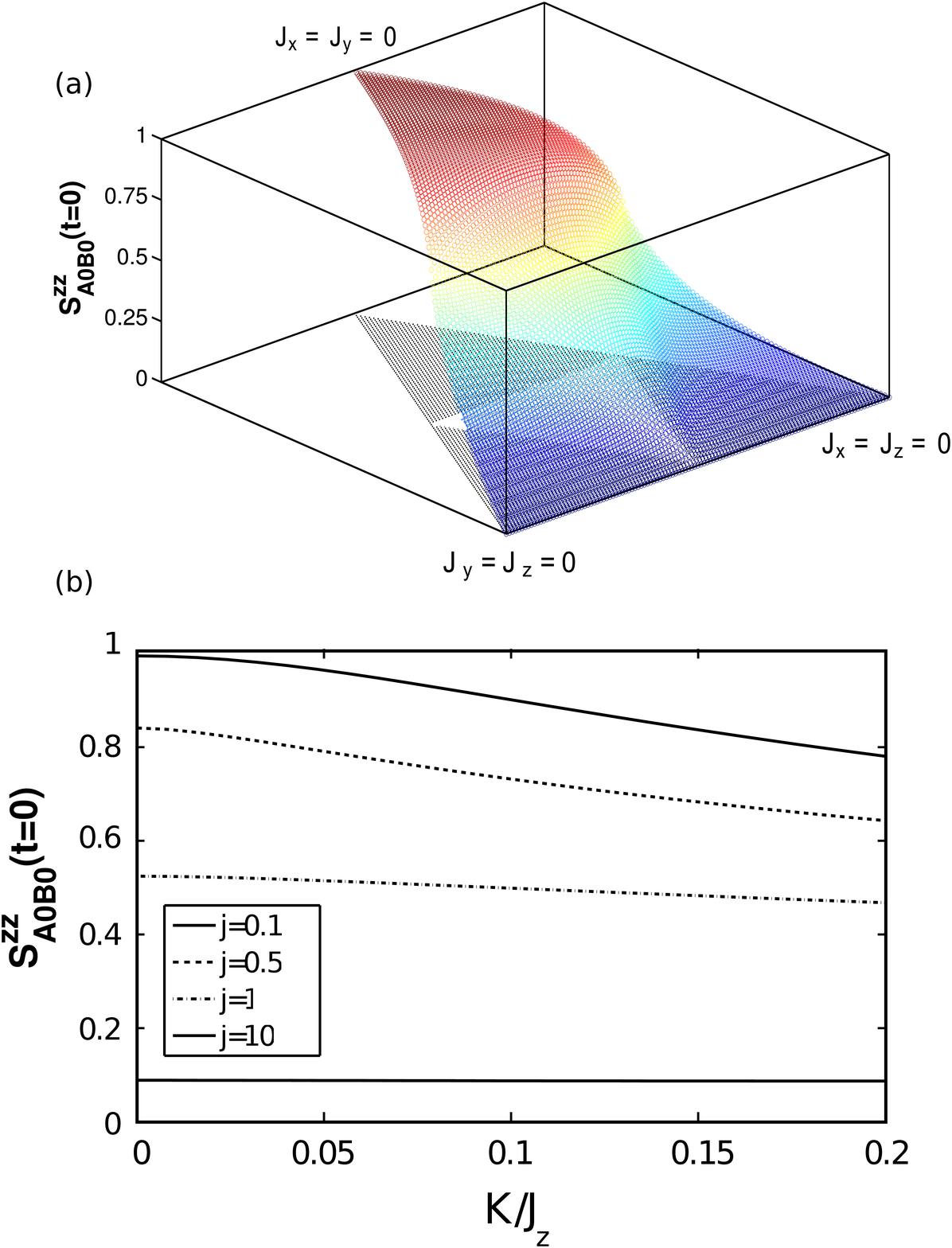}
\par\end{centering}
\caption{Dependence of the nearest-neighbour equal-time correlator $S^{zz}_{AB}(t=0)$ on the values of the exchange couplings. Panel (a) corresponds to the case $K=0$, where the strength of the n.n~correlator is shown along the cut $J_x+J_y+J_z=1$ in the parameter space. Panel (b), nearest-neighbour correlator as a function of $K/J_z$, at several fixed values of $j=J_x/J_z=J_y/J_z$, see inset.   
\label{EqualTimeSzz}}
\end{figure}

\subsection{Single particle contributions}\label{SectionSinglePartC}
In case (I), green region in Fig.~\ref{DynamicalPhaseDiagram}, the ground states of Majorana fermions have the same parity in both (initial/final) flux sectors. In order to calculate the spin-correlators from the Lehmann representation we insert the identity operator into Eq.~(\ref{SpinCorrFct1}) such that only the states of opposite parities contribute. Here we restrict the calculation to account for the contributions from single particle states. The latter have the form $|\lambda \rangle = \hat{b}_{\lambda}^{\dagger} |M_0^F\rangle$. With the help of the equation
\begin{equation}
\mathcal{X}_{\lambda j}-\mathcal{V}_{\lambda l} \mathcal{F}_{lj} = \left[ \mathcal{X}^{\dagger}\right]^{-1}_{\lambda j}
\end{equation}
the nearest-neighbour spin correlator assumes the form 
\begin{multline}
\label{SpinCorrFct1partFinalThreeSpin}
S^{zz(1)}_{AB} (t)  =  |\langle M_F|M_0 \rangle|^2 \sum_{\lambda}e^{it (E_0-E_{\lambda}^F)} \\ 
\times (X -Y)^{\dagger}_{0l} \mathcal{X}^{-1}_{l\lambda}(\mathcal{X}^{-1})^{\dagger}_{\lambda j}(X+Y)_{j0},
\end{multline}
where we use the summation convention on repeating indices. From this  expression (and a similar one for $S^{zz(1)}_{AA}$) the single particle contribution at $\mathbf{q}=0$ follows, Eq.~(\ref{StructureFactor1part}).

\section{Static correlators, and sum rules}\label{SumRules1}
As a check of our calculations of the dynamical correlation functions we make use of the sum rules, e.g.
\begin{equation}
\label{SumRules}
S^{zz} (t=0) =  \frac{1}{2\pi} \int_{-\infty}^{+\infty} d\omega\ S^{zz}(\omega),
\end{equation}
which relates equal time correlators to the integrated dynamical response. We employ this as a check of our exact results, as well as the quality of multi-particle approximation. The equal time correlator can be obtained from the equation
\begin{eqnarray}
\label{SpinCorrFctStatic}
S^{zz}_{AB} (t=0)  = \frac{1}{N} \sum_{\mathbf{q}\in \mathrm{BZ}} \cos \theta_{\mathbf{q}} 
\end{eqnarray}
with $\theta_\mathbf{q}$ defined in Eq.~(\ref{BogTrafo}). At the isotropic point for $K=0, J_a=1$ we obtain $S^{zz}_{AB} (t=0)=0.5249$ in the thermodynamic limit. In Fig.~\ref{EqualTimeSzz} (a) the equal time correlation function is shown in the full phase diagram~\cite{Baskaran2007} for $K=0$. The isolated Ising dimer limit $S^{zz}_{AB} (t=0)=1$ is quickly approached for $J_x,J_y\ll J_z, J_z=1$. In panel (b) of the same figure we show the evolution of $S^{zz}_{AB}(t=0)$ as a function of $K$ for different values of $j=J_x/J_z=J_y/J_z$. Static correlations always decrease with increasing $|K|$.
%

%%%%%%%%  

\end{document}